\RequirePackage{silence}
\documentclass[fleqn,usenatbib,useAMS]{mnras}
\usepackage{graphicx}
\usepackage{amsmath}
\usepackage{amsfonts}
\usepackage{float}
\usepackage{bm}
\usepackage[perpage,symbol*]{footmisc}
\usepackage{ae,aecompl}
\usepackage{array}
\usepackage{soul}
\usepackage{mathtools}
\usepackage{multirow}
\usepackage{pdflscape}

\newcommand{\appropto}{\mathrel{\vcenter{
  \offinterlineskip\halign{\hfil$##$\cr 
    \propto\cr\noalign{\kern2pt}\sim\cr\noalign{\kern-2pt}}}}}

\newcommand{\ssim}{\,{\sim}\,} 

\DeclareRobustCommand{\perthousand}{%
  \ifmmode
    \text{\textperthousand}%
  \else
    \textperthousand
  \fi}

\title[The Milky Way escape velocity curve in MOND]{The Escape Velocity Curve of the Milky Way in Modified Newtonian Dynamics} 

\author[Indranil Banik \& Hongsheng Zhao]{Indranil Banik$^{1}$\thanks{Email: \href{mailto:ib45@st-andrews.ac.uk}{ib45@st-andrews.ac.uk} (Indranil Banik)\newline $~~~~~~~~~~~~~~$ \href{mailto:hz4@st-andrews.ac.uk}{hz4@st-andrews.ac.uk} (Hongsheng Zhao)}, Hongsheng Zhao$^{1}$\\
$^{1}$Scottish Universities Physics Alliance, University of St Andrews, North Haugh, St Andrews, Fife, KY16 9SS, UK}

\pubyear{2017}
\begin{document}
\label{firstpage}
\pagerange{\pageref{firstpage}--\pageref{lastpage}}

\maketitle

\begin{abstract}


We determine the escape velocity from the Milky Way (MW) at a range of Galactocentric radii in the context of Modified Newtonian Dynamics (MOND). Due to its non-linear nature, escape is possible if the MW is considered embedded in a constant external gravitational field (EF) from distant objects. We model this situation using a fully self-consistent method based on a direct solution of the governing equations out to several thousand disk scale lengths. We try out a range of EF strengths and mass models for the MW in an attempt to match the escape velocity measurements of Williams et al. (2017).

A reasonable match is found if the EF on the MW is ${\ssim 0.03 a_{_0}}$, towards the higher end of the range considered. Our models include a hot gas corona surrounding the MW, but our results suggest that this should have a very low mass of ${\ssim 2 \times 10^{10} M_\odot}$ to avoid pushing the escape velocity too high. Our analysis favours a slightly lower baryonic disk mass than the ${\ssim 7 \times 10^{10} M_\odot}$ required to explain its rotation curve in MOND. However, given the uncertainties, MOND is consistent with both the locally measured amplitude of the MW rotation curve and its escape velocity over Galactocentric distances of 8$-$50 kpc.

\end{abstract}


\section{Introduction}
\label{Introduction}

The standard $\Lambda$ cold dark matter cosmological paradigm \citep[$\Lambda$CDM,][]{Ostriker_1995} has a great deal of flexibility in fitting the rotation curves (RCs) of individual galaxies due to the unknown relation between their baryonic content and their often dominant dark matter (DM) distribution required by this model \citep[e.g.][]{Rubin_1980}. This makes it difficult to extract unique RC predictions from $\Lambda$CDM, as illustrated by \citet{Blok_1998}. Their Figure 6 shows that $\Lambda$CDM can fit the RC of NGC 2403 rather well based on the photometry data of UGC 128, a different galaxy with a much lower surface brightness and indeed a rather different RC. In this model, it should therefore be difficult to predict the RCs of individual galaxies based solely on their baryonic distribution.

However, observations indicate the opposite \citep[][and references therein]{Famaey_McGaugh_2012}. In a $\Lambda$CDM context, spiral galaxy RCs exhibit a tight correlation between their shape, dark matter halo scale radius and mass such that $\la 10^{-5}$ of the available phase space volume is actually filled \citep{Salucci_2007}. Those authors noted that ``theories of the formation of spirals do not trivially imply the existence of such a surface that underlies the occurrence of a strong dark-luminous coupling''. Although it has long been known that the Newtonian gravity $\bm{g_{_N}}$ of the baryons is insufficient to \emph{explain} the RC of many galaxies, the RC can nonetheless be \emph{predicted} from $\bm{g_{_N}}$ alone by scaling it in a universally valid way. This radial acceleration relation (RAR) has recently been clarified with space-based Spitzer observations in the near-infrared \citep{McGaugh_Lelli_2016}, taking advantage of the lower dispersion in mass to light ratios at these wavelengths \citep{Bell_de_Jong_2001, Norris_2016}. Similar analyses are sometimes possible in elliptical galaxies, especially those that contain a thin rotation-supported gas disk \citep{den_Heijer_2015}. The RAR works extremely well in both classes of galaxy over ${\ssim 5}$ orders of magnitude in luminosity and a similar range of surface brightness \citep{Lelli_2017}. Deviations from the RAR fall within the expected observational uncertainties and appear to be uncorrelated with any of the numerous parameters that could plausibly be relevant e.g. surface brightness and gas fraction (see their Figure 4).

We consider the RAR to be more natural if galaxies are not surrounded by DM halos but their purported dynamical effect instead arises instead from an acceleration-dependent modification to Newtonian gravity, a hypothesis called Modified Newtonian Dynamics \citep[MOND,][]{Milgrom_1983}. This assumes that the gravitational field strength $g$ at distance $r$ from an isolated point mass $M$ transitions from the usual inverse square law at short range to
\begin{eqnarray}
	g ~=~ \frac{\sqrt{GMa_{_0}}}{r} ~~~\text{for } ~r \gg \sqrt{\frac{GM}{a_{_0}}} \, .
	\label{Deep_MOND_limit}
\end{eqnarray}

Here, $a_{_0}$ is a fundamental acceleration scale of nature which must have an empirical value close to $1.2 \times {10}^{-10}$ m/s$^2$ to match galaxy rotation curves \citep{McGaugh_2011}. In a remarkable coincidence called the cosmic coincidence of MOND, $a_{_0}$ is comparable to the value of $g$ at which a classical gravitational field has an energy density equal to the dark energy density $u_{_\Lambda} = \rho_{_\Lambda} c^2$ implied by the accelerating expansion of the Universe \citep{Riess_1998}. Thus,
\begin{eqnarray}
	\frac{g^2}{8\rm{\pi}G} ~<~ u_{_\Lambda} ~~\Leftrightarrow~~ g ~\la~ 2\rm{\pi}a_{_0} \, .
	\label{MOND_quantum_link}
\end{eqnarray}

This suggests that MOND may be caused by quantum gravity effects \citep[e.g.][]{Milgrom_1999, Pazy_2013, Verlinde_2016, Smolin_2017}. Regardless of its underlying microphysical explanation, it works well at explaining the dynamics of isolated galaxies. In the Local Group (LG), it requires that the Milky Way (MW) and M31 have undergone a past close flyby \citep{Zhao_2013} due to their strong gravitational attraction in MOND and the almost radial nature of the MW-M31 orbit \citep{Van_der_Marel_2012}. Such a flyby is not possible in $\Lambda$CDM because dynamical friction between their DM halos would inevitably cause a merger \citep{Privon_2013}. However, it would provide a natural explanation for several LG galaxies with very high radial velocities (RVs), much higher than can easily be accounted for in a 3D dynamical model of the LG in $\Lambda$CDM \citep{Banik_Zhao_2017}. These RVs remain anomalously high despite several improvements to the model and the procedure used to find its best-fitting parameters \citep[][section 4.2]{Banik_2017_anisotropy}. However, their figure 5 shows that the high-velocity galaxies (HVGs) have RVs broadly consistent with the speeds at which a once fast-moving MW or M31 could have slingshot out LG dwarfs in 3-body gravitational interactions governed by MOND.

One consequence of the MOND model should be that the HVGs lie rather close to the MW-M31 orbital plane. This is because it should be easier to scatter a dwarf galaxy to a very high RV if it is scattered parallel to the motion of the perturber. We recently used a test particle model to demonstrate this and also showed that the observed spatial distribution of the HVGs is indeed rather anisotropic \citep{Banik_2017_anisotropy}.

A past MW-M31 interaction might also have formed the thick disk of the MW \citep{Gilmore_1983}, a structure which formed fairly rapidly from its thin disk ${9 \pm 1}$ Gyr ago \citep{Quillen_2001}. More recent investigations confirm a fairly rapid formation timescale \citep{Hayden_2015} and an associated burst of star formation \citep[][figure 2]{Snaith_2014}. The disk heating which likely formed the Galactic thick disk appears to have been stronger in the outer parts of the MW, characteristic of a tidal effect \citep{Banik_2014}. This may be why the thick disk of the MW has a larger scale length than its thin disk \citep{Juric_2008, Jayaraman_2013}.

In this contribution, we test a more subtle consequence of MOND called the external field effect (EFE) which arises because the theory is acceleration-dependent \citep[][section 2g]{Milgrom_1986}. To understand it, consider a dwarf galaxy with low internal accelerations ($\ll a_{_0}$) freely falling in the strong acceleration ($\gg a_{_0}$) of a distant massive galaxy such that there are no tidal effects. The overall acceleration at any point in the dwarf is rather high due to the dominant external field (EF) of the massive galaxy. Thus, the dwarf would obey Newtonian dynamics and forces in its vicinity would follow the usual inverse square law rather than Equation \ref{Deep_MOND_limit}. However, without the massive galaxy, the internal dynamics of the dwarf would be very non-Newtonian.

Using the principle of continuity, the RC of a galaxy must be slightly affected even if the EF on it is much weaker than its internal gravity. Applying this idea, \citet{Haghi_2016} analysed whether the RCs of a sample of 18 disk galaxies could be fit better in MOND once the EFE is considered. Their work relied on a plausible analytic estimate of how the EFE would weaken the internal gravity of these galaxies. In most of the cases considered, non-zero values of the EF were preferred due to the RCs declining faster than expected in the outer regions if one neglects the EFE. Moreover, the preferred EF strengths were roughly consistent with the expected gravity from other known galaxies in the vicinity of the 18 they considered (see their Figure 7).

Perhaps the clearest demonstration of the EFE is in the velocity dispersion of the MW satellite Crater 2, which was predicted to be only $2.1^{+0.6}_{-0.3}$ km/s in MOND \citep[][section 3]{McGaugh_2016_Crater}. The rather low value is partly due to the EFE of the nearby MW, without which the prediction would have been ${\ssim 4}$ km/s. This is in tension with the observed value of $2.7 \pm 0.3$ km/s \citep{Caldwell_2017}. Thus, the internal dynamics of Crater 2 are not consistent with a naive application of the RAR but are consistent with a more rigorous treatment of MOND and its inevitable EFE.

At large distances from an object, the EF is likely to be the dominant source of gravity. We previously derived the far-field MOND forces generated by a point mass embedded in a constant and dominant EF of magnitude $g_{_{ext}}$ \citep{Banik_Zhao_2015}. $g$ eventually transitions to an inverse square law with a super-Newtonian normalisation if $g_{_{ext}} \ll a_{_0}$, as will be the case in this work. Thus, a point mass should create a potential well of finite depth even in MOND. Moreover, we showed that the escape velocity would differ by $\la 3\%$ between the most common versions of the theory, namely the aquadratic Lagrangian formulation \citep[AQUAL,][]{Bekenstein_Milgrom_1984} and the quasilinear formulation \citep[QUMOND,][]{QUMOND}. QUMOND is much simpler to handle numerically because it only requires a solution to the normal Poisson equation, albeit twice as often as for Newtonian gravity.\footnote{The increased computational cost is offset against the fact that QUMOND should work without the addition of dark matter particles, at least on galactic scales.} This is much simpler than the non-linear grid relaxation method required in AQUAL \citep{Brada_1999}, so we will focus on QUMOND in this contribution. This is also the basis for the publicly available Phantom of RAMSES algorithm \citep{PoR} which `MONDifies' the gravity law in the RAMSES $N$-body hydrodynamics solver \citep{Teyssier_2002}. A similar algorithm has recently been developed that can solve AQUAL as well, though this is not yet public \citep{Candlish_2015}.

Realising that MOND with the EFE predicts potential wells of finite depth, \citet{Famaey_2007} used an analytic method to estimate the escape velocity $v_{esc}$ from the MW in the vicinity of the Sun. Similar results were later obtained by \citet{Wu_2008} using a numerical solution to AQUAL. Their estimated $v_{esc}$ agrees reasonably well with later measurements based on high-velocity MW stars \citep{Piffl_2014}. Recently, a similar technique was used to measure $v_{esc}$ over a wide range of Galactocentric radii \citep[8$-$50 kpc,][]{Williams_2017}. This work applied the method of \citet{Leonard_1990} to a variety of tracers detected in the ninth data release of the Sloan Digital Sky Survey \citep[SDSS,][]{Ahn_2012}. We wish to calculate the expected $v_{esc}$ in MOND at these positions.

To better explore the range of plausible MW mass models, we include a hot gas halo surrounding its stellar and gas disks. This is suggested by X-ray spectroscopic observations at a range of Galactic latitudes \citep{Nicastro_2016} and by the truncation of the Large Magellanic Cloud gas disk, most likely a consequence of ram pressure stripping \citep{Salem_2015}. A similar halo has recently been detected around M31 based on quasar sightline observations \citep{Lehner_2015}.

In Section \ref{Method}, we explain the method by which we accurately determine $v_{esc}$ in our MOND models of the MW. The results thus obtained are shown in Section \ref{Results} and their accuracy is discussed in Section \ref{Discussion}, where we also consider other issues such as the plausibility of our best-fitting model parameters based on independent considerations. Our conclusions are given in Section \ref{Conclusions}.

\section{Method}
\label{Method}

QUMOND uses the Newtonian gravitational field $\bm{g_{_N}}$ due to a matter distribution as the first of two stages in calculating the gravitational field $\bm{g}$.
\begin{eqnarray}
	\label{QUMOND_governing_equation}
	\overbrace{\nabla \cdot \bm{g}}^{\propto \rho_{_{PDM}} + \rho_{_b}} &\equiv & \nabla \cdot \left[\nu \left( \frac{\left| \bm{g_{_N}} \right|}{a_{_0}}\right) \bm{g_{_N}} \right] ~~\text{, where} \\
      \nu \left( x \right) &=& \frac{1}{2} ~+~ \sqrt{\frac{1}{4} + \frac{1}{x}} \, .
\end{eqnarray}

Here, we use the `simple' interpolating function $\nu \left( x \right)$ to go between the Newtonian and low-acceleration regimes \citep{Famaey_Binney_2005}. $\nabla \cdot \bm{g}$ is the source term for the gravitational field, so it can be thought of as an `effective' density $\rho$ composed of the actual density $\rho_{_b}$ and an extra term which we define to be the phantom dark matter density $\rho_{_{PDM}}$. This is the distribution of dark matter that would be necessary in Newtonian gravity to generate the same gravitational field as in QUMOND. The Newtonian gravitational field $\bm{g_{_N}} \equiv -\nabla \Phi_N$ satisfies the usual Poisson equation
\begin{eqnarray}
	\nabla^2 \Phi_N ~=~ 4\pi G \rho_{_b} \, .
	\label{Poisson_equation}
\end{eqnarray}

After solving this with the usual isolated boundary conditions ($g \to 0$ as $r \to \infty$), we add the contribution from the Newtonian EF $\bm{{g}_{N,ext}}$ which is what the EF on the MW would have been in Newtonian gravity. We assume the spherically symmetric MOND relation between this and the actual EF $\bm{{g}_{ext}}$ on the MW.
\begin{eqnarray}
	\label{External_field_increment}
	\Phi ~\to~ \Phi - \bm{r} \cdot \bm{{g}_{N,ext}} &~& \text{, where} \\
	\overbrace{\nu \left(\frac{{g}_{_{N,ext}}}{a_{_0}} \right)}^{\nu_{_{ext}}} \bm{{g}_{N,ext}} ~&=&~ \bm{g}_{ext} \, .
\end{eqnarray}

\subsection{The Newtonian potential}
\label{Newtonian_gravity}

The MW is assumed to consist of a hot gas corona surrounding two aligned and concentric infinitely thin exponential disks representing its gas and stellar components. Taking advantage of the fact that we can superpose potentials in Newtonian gravity, we simply add the potential of the corona to that of the other components. The corona is treated as a Plummer model \citep{Plummer_1911} with mass $M_{cor}$ and core radius $r_{_{cor}}$, yielding a corona potential at distance $r$ from the MW of
\begin{eqnarray}
	\Phi_{cor} ~=~ -~\frac{GM_{cor}}{\sqrt{r^2 + {r_{_{cor}}}^2}} \, .
	\label{Corona_potential}
\end{eqnarray}

For the disk components, we only need to solve for a single exponential disk. We take this to have unit scale length and $GM$ so that it can be scaled up to the required values later. We discretise Equation \ref{Poisson_equation} and solve it in spherical polar co-ordinates (polar angle $\theta$) using the successive over-relaxation method described in Appendix \ref{Appendix_A}.  


Once we obtain $\Phi$ in this way, we scale it up to the correct mass and length scale for the MW stellar disk and superpose another scaled version of this solution to represent its gas disk. We then add the corona potential (Equation \ref{Corona_potential}) and the external field (Equation \ref{External_field_increment}).

\subsection{The QUMOND gravitational field}
\label{QUMOND_gravity}

Using the discretisation scheme described in Appendix \ref{Appendix_B}, we determine the QUMOND source density $\nabla \cdot \bm{g} \equiv \nabla \cdot \left( \nu \bm{g_{_N}} \right)$. We integrate this directly in order to obtain $\bm{g}$ itself, which we only need at a small fraction of the grid points.
\begin{eqnarray}
	\bm{g \left( \bm{r} \right)} ~=~ \int \nabla \cdot \bm{g} \left( \bm{r'}\right) \frac{\left( \bm{r} - \bm{r'} \right)}{|\bm{r} - \bm{r'}|^3} ~d^3\bm{r'} \, .
\end{eqnarray}

For axisymmetric problems, the forcing $\nabla \cdot \left( \nu \bm{g_{_N}} \right)$ can be considered as a large number of uniform density rings, making it simple to determine $\bm{g}$ on the symmetry axis via direct summation. For off-axis points, we avoid an excessive computational cost by making use of a `ring library' which stores the gravity exerted by a thin ring with $GM_{ring} = r_{_{ring}} = 1$, where $M_{ring}$ and $r_{_{ring}}$ refer to the ring mass and radius, respectively. To find the gravity at any point due to a ring, we scale the relative co-ordinates to the appropriate position within our ring library and interpolate to obtain the required result before scaling by $\frac{GM_{ring}}{{r_{_{ring}}}^2}$ at the end. We neglect rings passing very close to the point on which we calculate $\bm{g}$ as these rather large contributions should almost completely cancel, but it is difficult to handle such cancellation accurately on a computer. We consider the effect of slightly different `excluded regions' and use cubic extrapolation to estimate $\bm{g}$ if there had been no excluded region at all.

We can only consider contributions to $\bm{g}$ from a finite volume. Our aim is to consider a sufficiently large volume that contributions from more distant regions can be handled analytically. In particular, we cover a large enough region that the EF should start dominating beyond it. This transition occurs at a distance of
\begin{eqnarray}
	r_{_{ext}} ~=~ \frac{\sqrt{GMa_{_0}}}{g_{ext}} \, .
	\label{r_ext}
\end{eqnarray}

In the models we consider (Table \ref{Parameters}), $r_{_{ext}}$ can be as distant as 1220 kpc. Thus, we use our method to obtain $\nabla \cdot \bm{g}$ out to a distance $r_{out} = 13,341$ kpc. The phantom dark matter in the EF-dominated region beyond this is not spherically symmetric and thus contributes to the gravity at smaller distances. We account for this assuming EF dominance in the regions beyond $r_{out}$, allowing us to analytically determine the gravity resulting from phantom dark matter there as that follows the density distribution derived in \citet[][section 3]{Banik_Zhao_2015}. This leads to an adjustment to $\bm{g}$ of
\begin{eqnarray}
	\label{Exterior_PDM_adjustment}
	\Delta \bm{g} ~&=&~ \frac{2GM\nu_{_{ext}}K_0}{15 \, {r_{out}}^3} \left( \bm{r} - 3 r \cos \theta \, \widehat{\bm{g}}_{_{ext}} \right) \text{, where } \nonumber \\ 
    \widehat{\bm{g}}_{_{ext}} ~&\equiv&~ \frac{\bm{g}_{_{ext}}}{\left| \bm{g}_{_{ext}}\right|} \, .
\end{eqnarray}

Assuming that we are considering a point where the EF is dominant, the gravity due to the MW has a magnitude of $\ssim \frac{GM\nu_{_{ext}}}{r^2}$ such that the correction to it expressed in fractional terms is only $\ssim \frac{1}{15} \left(\frac{r}{r_{out}} \right)^3$. Thus, the accuracy of our results should not depend much on this correction, which should in any case be fairly accurate as it estimates contributions from regions with $r > 11r_{_{ext}}$. There, the Newtonian gravitational field due to the MW should be $> 120 \times$ weaker than $g_{_{N,ext}}$, allowing it to be considered perturbatively.

To minimise edge effects of the sort just mentioned, we only use our method to obtain $\bm{g}$ out to a distance of 3892 kpc where Equation \ref{Exterior_PDM_adjustment} requires us to correct $\bm{g}$ by ${\ssim \frac{1}{600}}$. The EF is not really dominant at this point, so some method is required to estimate $\bm{g}$ at larger distances. Only then can we determine $\bm{g}$ far enough out into the EF-dominated region for us to `hand over' to the analytic results found by \citet{Banik_Zhao_2015}. For this purpose, we construct another library holding the gravitational field due to a point mass embedded in a constant EF assuming the deep-MOND limit (DML)\footnote{to avoid recalculating it for different EF strengths}. It is not totally accurate to assume the DML because the EF is still a few percent of $a_{_0}$ \citep{Famaey_2007}. Thus, we adjust forces in this region by the ratio of $\nu_{_{ext}}$ to its value $\nu_{_{DML}}$ assuming the DML.
\begin{eqnarray}
	\frac{\nu}{\nu_{_{DML}}} ~&=&~ 1 ~+~ \frac{g_{_{ext}}}{a_{_0}} ~~\text{, where} \\
	\nu_{_{DML}} ~&\equiv&~ \sqrt{\frac{a_{_0}}{g_{_{N,ext}}}} \, .
	\label{nu_adjustment}
\end{eqnarray}

To see how appropriate a point mass model is for the MW, we need to consider its most extended component $-$ its hot gas corona. The fraction of its total mass enclosed within any radius $r$ is given by
\begin{eqnarray}
	\frac{M \left( < r \right)}{M} ~=~ \frac{r^3}{\left( r^2 + {r_{_{cor}}}^2\right)^\frac{3}{2}} \, .
	\label{Plummer_distribution}
\end{eqnarray}

In the most extended case we consider, $r_{_{cor}} = 60$ kpc so that only $\ssim 10^{-4}$ of the corona lies at radii beyond 3.9 Mpc. Having thus verified the accuracy of our point mass and DML assumptions and corrected their deficiencies as far as possible, we use them to obtain the gravitational field out to $r = 66.5r_{_{ext}}$. Here, we find that the forces are within a few percent of the analytic results of \citet{Banik_Zhao_2015} which we therefore use to obtain the depth of the potential well at this point.
\begin{eqnarray}
	\label{Phi_g_ext_domination}
	\Phi &=& ~-\frac{GM\nu_{_{ext}}}{r} \left( 1 + \frac{K_{_0}}{2} \sin^2 \theta \right) ~\text{, where} \\
	K_0 &\equiv & \frac{\partial Ln ~ \nu_{_{ext}}}{\partial Ln ~ {g}_{_{N, ext}}} = ~-\frac{1}{2}~~\text{ if } {g}_{_{N, ext}} \ll a_{_0} \, .
\end{eqnarray}

To summarise, our potential calculations are based on considering the gravitational field in three regions. The innermost one covers out to 3892 kpc and considers the MW mass distribution in detail as well as the EF it is embedded in, without assuming anything about how strong the resulting gravity is. In the next region covering out to $66.5r_{_{ext}}$, the MW is treated as a point mass in an EF without assuming the gravity from either is dominant. Although this aspect of the problem is assumed to be in the DML, a small correction is applied to account for $g_{_{ext}}$ being a few percent of $a_{_0}$. The outermost region uses the analytic potential arising from a point mass embedded in a dominant EF (Equation \ref{Phi_g_ext_domination}).

\section{Results}
\label{Results}

\begin{table}
 \begin{tabular}{lll}
	\hline
  Variable & Meaning & Value \\ 
  \hline
  $R_\odot$ & Galactocentric distance of Sun & 8.2 kpc\\
  $r_*$ & Stellar disk scale length & 2.15 kpc\\
  $M_{*,0}$ & Nominal stellar disk mass & $5.51 \times 10^{10} M_\odot$\\
  $r_{_g}$ & Gas disk scale length & 7 kpc\\
  $M_{g,0}$ & Nominal gas disk mass & $1.18 \times 10^{10} M_\odot$\\ \hline
	$\frac{M_*}{M_{*,0}}$ & Disk mass scaling factor & $0.8-1.4$\\ 
  $r_{_{cor}}$ & Plummer radius of corona & $\left(20-60 \right)$ kpc\\
  $M_{_{cor}}$ & Corona mass & $\left(2 - 8 \right)\times 10^{10} M_\odot$\\
  $g_{_{ext}}$ & External field on MW & $\left(0.01 - 0.03\right)a_{_0}$\\
  \hline
 \end{tabular} 
 \caption{Our adopted parameters for the MW mass distribution, with a $_0$ subscript indicating a nominal value. We always use the same value of $\frac{M_g}{M_*}$ and the same disk scale lengths, but vary the other parameters using a grid search. The first part of our table contains the fixed parameters $R_\odot$ \citep{McMillan_2017}, $M_{*,0}$ \citep{McGaugh_2016_MW}, $r_*$ \citep{Bovy_2013}, $r_{_g}$ \citep{McMillan_2017} and $M_{g,0}$, which is based on applying the method described in \citet[][section 3.3]{McGaugh_2008} to the observations of \citet[][table D1]{Olling_Merrifield_2001}. The Galactic hot gas corona is modelled using Equation \ref{Plummer_distribution} \citep{Plummer_1911}.}
 \label{Parameters}
\end{table}

Our model for the MW mass distribution is designed to be consistent with its observed rotation curve in a MOND context \citep[][table 1 model Q4ZB]{McGaugh_2016_MW}. This particular model does not require a bulge to get the gravitational field strength correct in the Solar neighbourhood because of its large distance from the Galactic centre and the rather short length scale associated with any bulge component. Indeed, \citet{McMillan_2017} suggest that 80\% of its mass lies within just 2.2 kpc (see their section 2.1) whereas the Sun is nearly $4 \times$ more distant (see their Table 2). At even larger distances, the dynamical effect of the bulge should be even smaller and easily accounted for with minor adjustments to the disk normalisations and scale lengths. Moreover, their work also contains a central hole in the MW gas disk whereas our model does not, partly compensating for our lack of a bulge component. Thus, our results should not be greatly affected by this assumption.

\begin{figure}
	\centering 
	\includegraphics [width = 8.5cm] {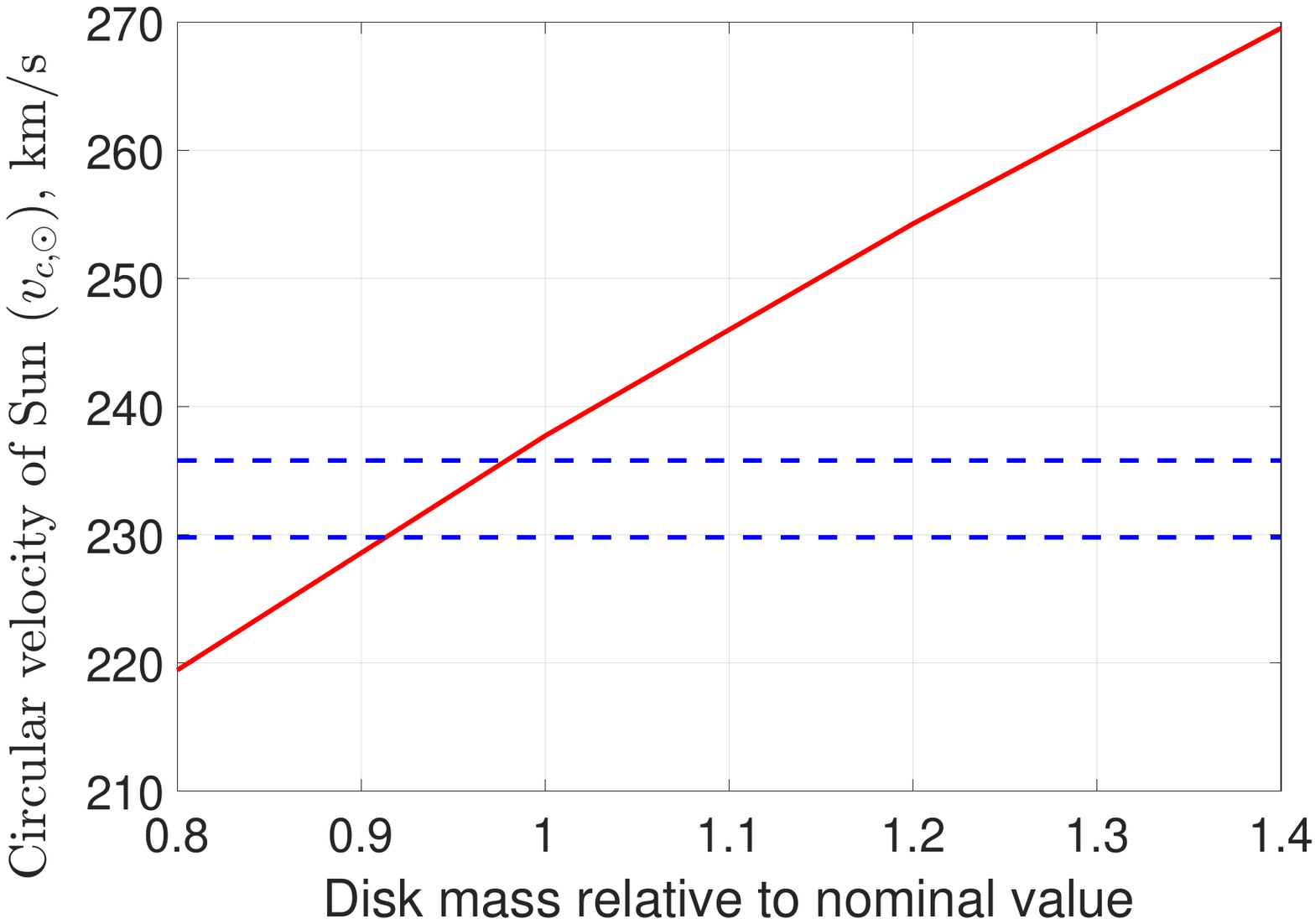}
	\caption{The effect on $v_{c, \odot}$ of scaling our nominal stellar and gas disk masses (Table \ref{Parameters}) by the amount shown. The dashed blue lines show the 1$\sigma$ allowed range for $v_{c, \odot}$ \citep{McMillan_2017}.}
	\label{LSR_velocity}
\end{figure}

We begin by considering how our 4 model parameters (Table \ref{Parameters}) influence $v_{c, \odot}$. This depends mainly on the disk surface density such that only the nominal model is able to reproduce this correctly if we assume that $v_{c, \odot} \approx 235$ km/s \citep{McMillan_2017}. However, we consider the effect of scaling the surface density by factors of 0.8$-$1.4 (Figure \ref{LSR_velocity}).

Adjusting $M_{_{cor}}$ can affect $v_{c, \odot}$ by $\la 5$ km/s while adjusting its scale length $r_{_{cor}}$ has a smaller effect of ${\ssim 1}$ km/s. At the Solar circle, the MW is effectively isolated $-$ raising $g_{_{ext}}$ from the lowest to the highest values considered only affects $v_{c, \odot}$ by $\ssim 4$ m/s. These factors must have a more significant influence on forces further from the MW, but the scarcity of tracers makes it difficult to directly measure $\bm{g}$ there. Fortunately, forces at large $r$ affect the escape velocity $v_{esc} = \sqrt{-2\Phi}$ near the Sun, which must therefore be rather more sensitive to these parameters. This gives us the opportunity to constrain the MW gravitational field at large distances based on observations relatively close to it.

\begin{figure}
	\centering 
	\includegraphics [width = 8.5cm] {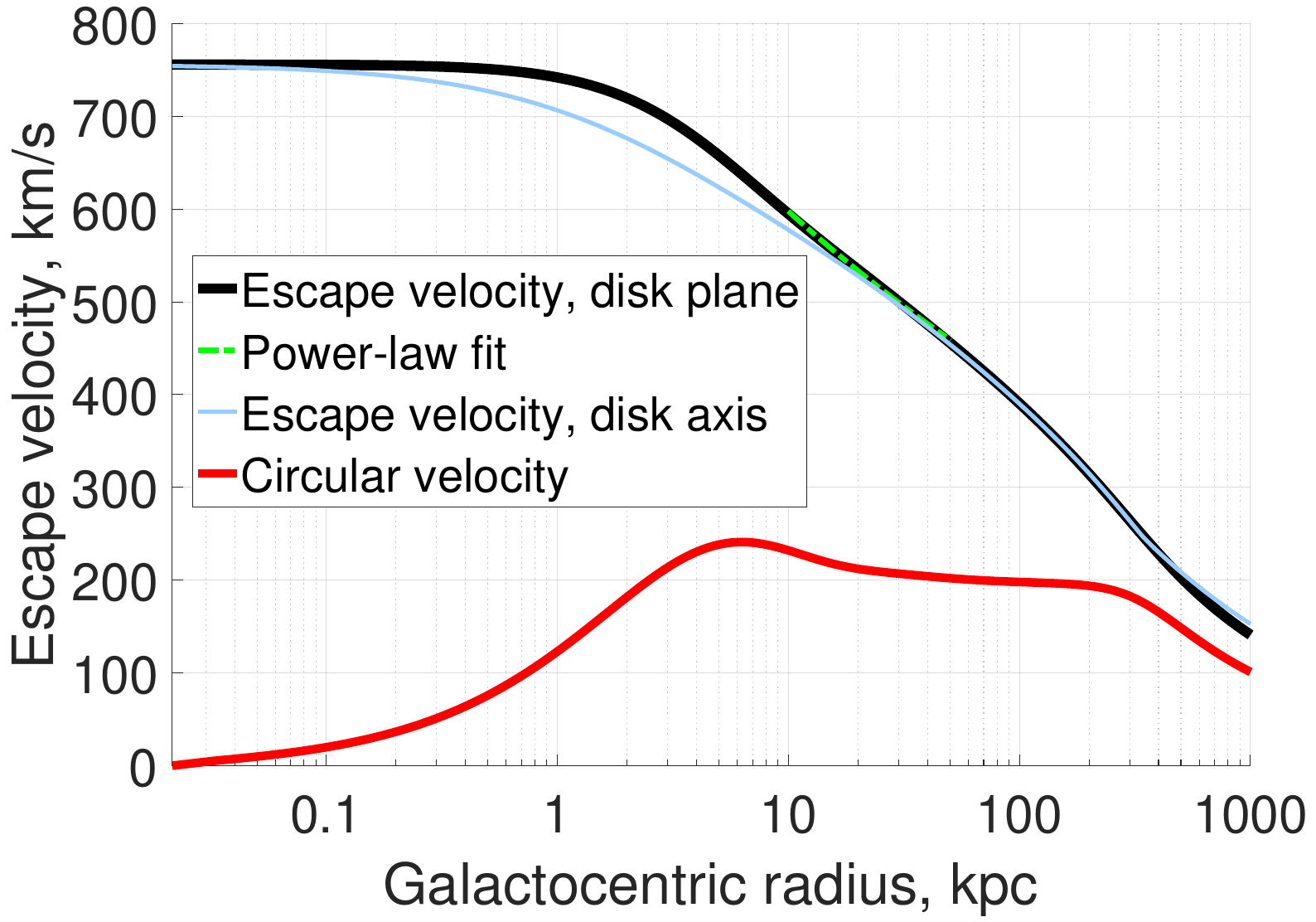}
	\caption{How the circular velocity of the MW (lower red curve) and its escape velocity (upper black curve) depend on position within its disk plane. The latter can be parameterised rather well as a power law (Equation \ref{Power_law_fit}) over the radial range $10-50$ kpc (dashed green curve). At the same distance from the MW, its escape velocity is lower along its disk axis (thin blue curve) for points close to the MW due to the effect of its disk. However, this pattern is reversed at long range because we assume the EF on the MW is aligned with its disk axis, deepening the potential in this direction \citep[][equation 37]{Banik_Zhao_2015}. The model shown here uses the nominal disk masses in Table \ref{Parameters} and $g_{_{ext}} = 0.03a_{_0}$, with the corona being as small and low-mass as possible.}
	\label{v_esc_profile}
\end{figure}

Using our nominal values for the MW stellar and gas disk masses (Table \ref{Parameters}), we determined its circular and escape velocity curves within its disk plane (Figure \ref{v_esc_profile}). For comparison with observations, we fit a power-law model to $v_{esc}$ over the radial range 10$-$50 kpc. This assumes that
\begin{eqnarray}
	v_{esc} \left( r \right) ~\propto~ r^{-\alpha} \, .
	\label{Power_law_fit}
\end{eqnarray}

Power-law fits become linear when considering the logarithms of both variables. Thus, if we let $y \equiv Ln~v_{esc}$ and $x \equiv Ln~r$ be lists of size $N$, then we have that
\begin{eqnarray}
	\tilde{x} ~&\equiv&~ x - \frac{1}{N}\sum_{i = 1}^N x_{_i} ~~~\left(\tilde{y}~\text{defined analogously} \right), \\
	\alpha ~&=&~ -\frac{\sum_{i = 1}^N \tilde{x}_{_i} \tilde{y}_{_i} }{\sum_{i = 1}^N \tilde{x}_{_i} \tilde{x}_{_i}} \, .
	\label{alpha_definition}
\end{eqnarray}

\begin{figure}
	\includegraphics [width = 12.5cm] {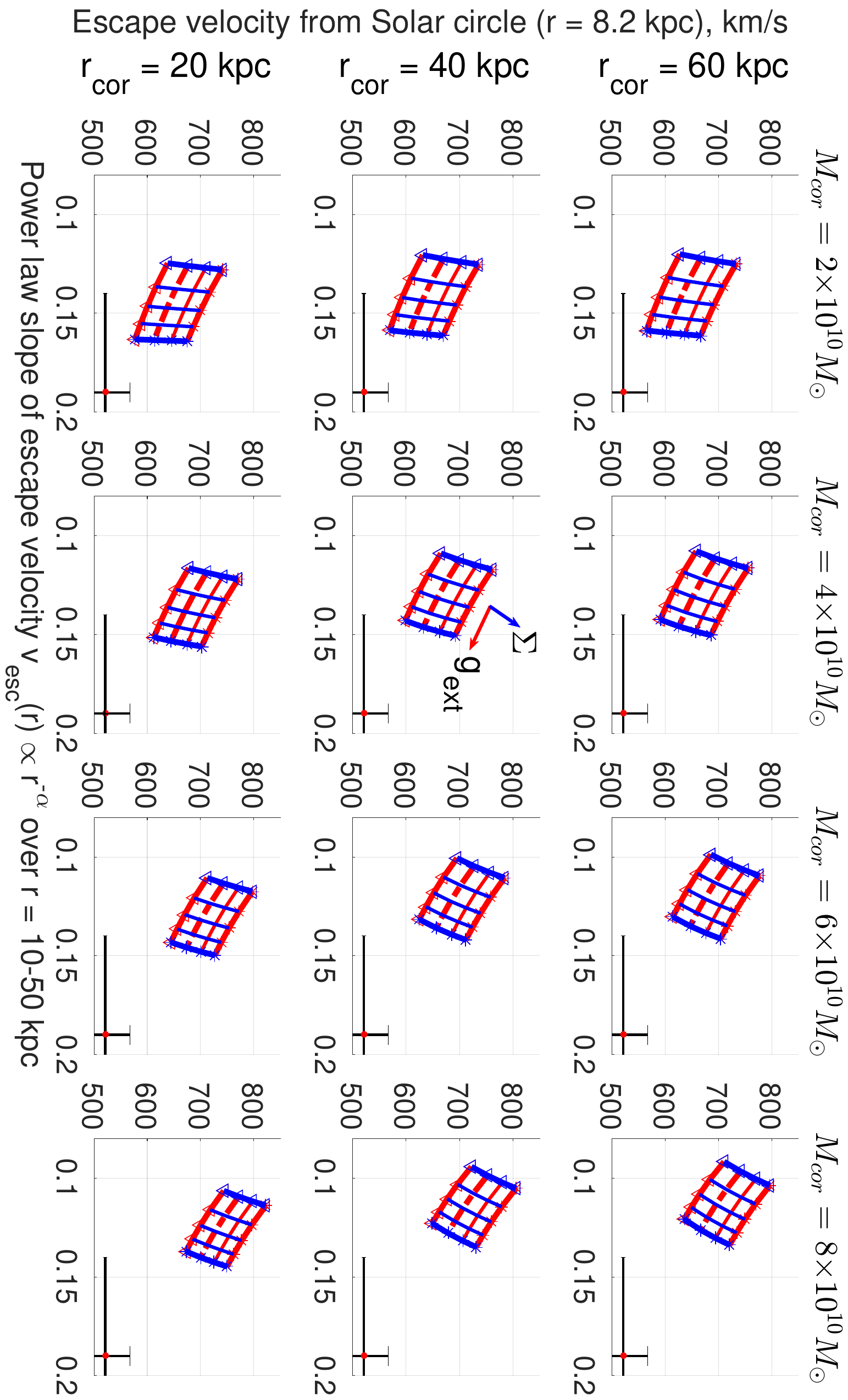}
	\caption{Escape velocity $v_{esc}$ from within the MW disk plane as a function of the model parameters. The $x$-axis shows the value of $\alpha$ such that $v_{esc} \left( r \right) ~\propto~ r^{-\alpha}$ while the $y$-axis shows $v_{esc}$ near the Sun. The measured values of these quantities are shown as a red dot with black error bars towards the bottom right \citep{Williams_2017}. Each subplot has a fixed corona mass and scale length, with red tracks showing the effect of varying $g_{_{ext}}$ with constant disk mass (vice versa for blue tracks). In each case, an inverted triangle is used to show the result when the parameter being varied has the lowest value we consider while a star is used for the largest value. We consider disk masses scaled from their nominal values by factors given in Table \ref{Parameters}, where we also show the range in $g_{_{ext}}$ that we try (values of all parameters are spaced linearly). The dashed red lines show the results for the nominal stellar and gas disk masses, which is required to obtain the correct $v_{c, \odot}$ (Figure \ref{LSR_velocity}). We assume $\bm{g_{_{ext}}}$ is aligned with the disk symmetry axis. Rotate $90^\circ$ anti-clockwise for viewing.}
	\label{v_esc_disk_results}
\end{figure}

\begin{figure}
	\includegraphics [width = 15.5cm] {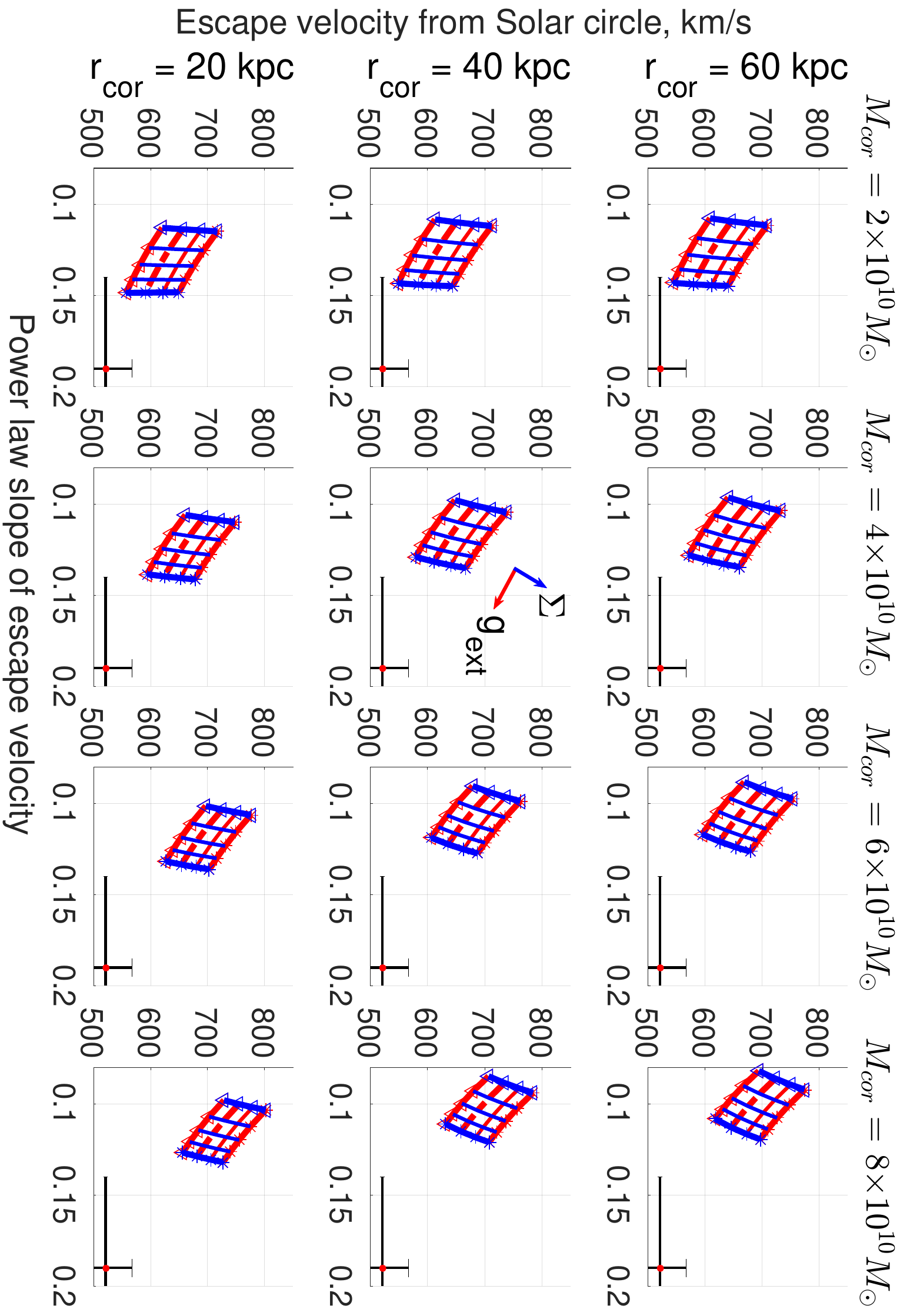}
	\caption{Similar to Figure \ref{v_esc_disk_results} but for $v_{esc}$ along the disk symmetry axis in the direction of $\bm{g_{_{ext}}}$. We obtained very similar results when going in the opposite direction (not shown), presumably because the radial gravity at both short and long range (isolated or EF-dominated) is the same at points with polar angles of $\theta$ and $\pi - \theta$. Rotate $90^\circ$ anti-clockwise for viewing.}
	\label{v_esc_results_top_axis}
\end{figure}

We can fit $v_{esc} \left( r \right)$ rather well using a power law over the range $r = 10-50$ kpc (Figure \ref{v_esc_profile}). For a range of models, we compare the resulting slope $\alpha$ and normalisation at the Solar circle to observations (Figures \ref{v_esc_disk_results} and \ref{v_esc_results_top_axis}). It is unclear exactly which Galactic polar angles $\theta$ the observations of \citet{Williams_2017} correspond to, but most likely a range of angles is used in order to get enough of the relatively rare high-velocity stars that are necessary for an escape velocity determination. Thus, we show results for points along the disk symmetry axis ($\theta = 0$) in Figure \ref{v_esc_disk_results} and within the disk plane ($\theta = \frac{\pi}{2}$) in Figure \ref{v_esc_results_top_axis}. The escape velocities are slightly larger in the latter case because the MW matter distribution gets closer to a point within its disk plane than an equally distant point along its disk axis. Within ${\ssim 100}$ kpc of the MW, this near-field effect is more important than the non-sphericity of the MW potential in the far-field EF-dominated region (Equation \ref{Phi_g_ext_domination}) where the MW exerts very little gravity in any case. However, beyond ${\ssim 100}$ kpc, the latter effect dominates because the MW can be considered as a point mass. This leads to a deeper potential in the direction of $\bm{g_{_{ext}}}$, which we take to be aligned with the disk axis for simplicity (Figure \ref{v_esc_profile}).

The results for $\theta = 0$ are hardly affected if we set $\bm{g_{_{ext}}} \to -\bm{g_{_{ext}}}$ because the forces at long range are symmetric with respect to $\theta \to \pi - \theta$ (Equation \ref{Phi_g_ext_domination}). Naturally, they are also symmetric at short range as the EF is unimportant here and the MW model is symmetric. This minimises any scope for $v_{esc}$ differing between equally distant points along $\theta = 0$ and $\pi$, or equivalently at the same point but with $\bm{g_{_{ext}}} \to -\bm{g_{_{ext}}}$. Although our algorithm is unable to rigorously consider intermediate EF orientations, this should have only a small effect on our results (Section \ref{External_field_direction}).


\onecolumn

\twocolumn

\section{Discussion}
\label{Discussion}

Our calculations for the MW's escape velocity are broadly consistent with the observations of \citet{Williams_2017} over the Galactocentric distances they cover ($8-50$ kpc), especially when considering that the Solar neighbourhood escape velocity could be as high as 690 km/s at the 95\% confidence level (see their Figure 3). Moreover, it is probably easier to underestimate $v_{esc}$ than to overestimate it. The velocity distribution of stars is expected to drop off close to $v_{esc}$, but it is easy to imagine the theoretical distribution function not being filled all the way up to $v_{esc}$ because it is impossible to have less than one star. This could well be why earlier studies underestimated $v_{esc}$ locally. For instance, \citet{Meillon_1998} found a 90\% confidence upper limit of just 550 km/s despite using Hipparcos data \citep{Perryman_1989} supplemented by accurate radial velocities \citep{Udry_1997}. Even if the positions and velocities of all MW stars were known perfectly, we could quite plausibly find none moving faster than $0.9 v_{esc}$ \citep[][figure 1]{Smith_2007}.

This issue should have been alleviated somewhat by the efforts of \citet{Williams_2017} to constrain the precise form of the cut-off in the velocity distribution close to $v{_{esc}}$ (see their section 5.1). Nonetheless, it must persist at some level and likely becomes more severe further from the MW due to the lower stellar number density. This can lead to a faster apparent decline in $v_{esc}$ than is actually the case, perhaps explaining their rather high inferred value of $\alpha$ (Equation \ref{Power_law_fit}) compared to our models.

Another issue might be that stars do not need to get infinitely far from the MW in order to escape. If they get to $400$ kpc, then this is half-way to M31 in some directions \citep{McConnachie_2012}. At this point, our calculations suggest that $v_{{esc}}$ is still $\ssim 150$ km/s (Figure \ref{v_esc_profile}). Subtracting this from 600 km/s in quadrature suggests that the local escape velocity could be reduced by $\ssim 20$ km/s due to the presence of M31. This effect would be larger 50 kpc from the MW, where $v_{esc}$ is only $\ssim 450$ km/s. Thus, $v_{esc} \left(r \right)$ might well decline faster than in our models. This could raise $\alpha$ by $\ssim 0.02$, possibly more if M31 is heavier than the MW, as appears likely \citep{Jorge_2014, Banik_Zhao_2016}. Such effects would depend on the direction relative to the direction towards M31, necessitating a fully 3D model for the MW because M31 does not lie very close to the Galactic disk plane. In this case, it would be important to carefully consider the survey volume to better understand how M31 might influence $v_{esc}$. This is beyond the scope of our analysis and probably not worthwhile given the present uncertainties on $v_{esc}$, but may become important in future when more accurate measurements become available.

One indication that these effects do not greatly influence the analysis of \citet{Williams_2017} comes from a relation between the circular and escape velocity curves that arises because both are determined by the same potential $\Phi$.
\begin{eqnarray}
	\frac{\partial \left( \frac{1}{2} {v_{esc}}^2 \right)}{\partial r} ~&=&~ -\frac{\partial \Phi}{\partial r} \\
	&=&~ -\frac{{v_{c}}^2}{r}  \, , \\
	\alpha ~\equiv~ -\frac{\partial Ln~v_{esc}}{\partial Ln~r} ~&=&~ \left( \frac{v_{c}}{v_{esc}}\right)^2 \, .
\end{eqnarray}

If we assume that $v_{c, \odot} = 232.8$ km/s \citep{McMillan_2017}, then we expect $\alpha = 0.200$ for a local escape velocity of 521 km/s. This is entirely consistent with the observed value of $0.19 \pm 0.05$. The mild tension could indicate that $\alpha$ is indeed 0.2 rather than 0.19, though our analysis suggests that the solution lies instead with higher $v_{esc}$.

Our models treat the disk components of the MW as infinitely thin. This seems a reasonable approximation given the rather small scale heights of the MW thin and thick disks \citep{Snaith_2014}. Still, the disk components have a finite thickness, yielding a shallower potential well within the disk plane. This would bring our model more in line with observations if they are primarily sensitive to low Galactic latitudes (Figure \ref{v_esc_disk_results}). If instead they are more sensitive towards the Galactic poles, then the effect of thickening the disk is likely smaller. This is fortunate as our model predictions are already more consistent with observations if they are along the disk axis (Figure \ref{v_esc_results_top_axis}).

Our results are somewhat uncertain due to imperfect knowledge of $R_\odot$. Presently, this is constrained to within $\ssim 0.1$ kpc \citep{Chatzopoulos_2015, McMillan_2017}. Increasing $R_\odot$ from our adopted 8.2 kpc to 8.5 kpc reduces $v_c$ and $v_{esc}$ in the Solar neighbourhood by $\ssim 1$ and 3 km/s, respectively. These effects are much smaller than the observational uncertainties, especially for $v_{esc}$.

The uncertain LSR speed $v_{c, \odot}$ also has some effect on our results. The analysis of \citet{McMillan_2017} found that $v_{c, \odot}$ could be as low as 220 km/s if the present constraints on $R_\odot$ are not considered. In this case, MOND requires a lower surface density $\Sigma$ for the MW to match a slower rotation curve. Figure \ref{LSR_velocity} suggests that we might need to scale $\Sigma$ by $\ssim 0.8$, thereby reducing $v_{esc}$ near the Solar neighbourhood by $\ssim 40$ km/s while leaving $\alpha$ almost unaltered (Figure \ref{v_esc_disk_results}).

However, such models rely on a very low value of $v_{c, \odot}$ which in turn implies a rather low $R_\odot$ given tight constraints on the ratio of these quantities from the proper motion of the supermassive black hole at the centre of the MW \citep{Brunthaler_2007}. This is in significant tension with independent measurements of $R_\odot$. Once these are considered, it becomes clear that $v_{c, \odot}$ is constrained to be $232.8 \pm 3.0$ km/s, making a value of 220 km/s highly unlikely \citep{McMillan_2017}. It is thus difficult to improve our results significantly through tighter constraints on the position and velocity of the Sun with respect to the MW. Instead, we suggest that observers should focus on improving measurements of $v_{esc}$.

Our results shed new light on the issue of whether the Large Magellanic Cloud (LMC) is bound to the MW. At a Galactocentric distance of 50 kpc \citep{Pietrzynski_2013}, we expect that $v_{esc} \approx 440$ km/s for a MW disk having the mass required to explain its circular velocity curve in a MOND context \citep[][table 1 model Q4ZB]{McGaugh_2016_MW}. If we scale the disk mass down by 0.8 (lowest red tracks in Figures \ref{v_esc_disk_results} and \ref{v_esc_results_top_axis}), then $v_{esc}$ would fall by $\approx 30$ km/s. Even then, 410 km/s greatly exceeds the Galactocentric velocity of the LMC as this is only ${321 \pm 24}$ km/s \citep[][table 5]{Kallivayalil_2013}. Thus, the LMC is almost certainly a bound satellite of the MW in a MOND context unless the EF on it significantly exceeds the maximum value of 0.03$a_{_0}$ considered here, a conclusion also reached by \citet{Wu_2008}. Even without considering dynamical models of the MW, the observations of \citet{Williams_2017} alone indicate that $v_{esc} = 379^{+34}_{-28}$ km/s at the distance of the LMC, strongly suggesting that it is bound to the MW.

Although our analysis is based on standard MOND, this tends to underestimate the forces binding galaxy clusters \citep[e.g.][]{Sanders_2003}. One possible solution is Extended MOND \citep[EMOND,][]{Zhao_2012} which posits that the acceleration scale $a_{_0}$ increases with the potential depth $\left| \Phi \right|$ such that it has the standard value of $1.2 \times {10}^{-10}$ m/s$^2$ for $\left| \Phi \right| \ll \Phi_0$ but is larger in regions with a deeper potential, such as galaxy clusters. So far, this appears to be a promising way to resolve the difficulties typically faced by MOND in such systems \citep{Hodson_2017}. The required value of $\Phi_0$ corresponds to a speed of $\sqrt{2 \Phi_0} =$ 1800 km/s (see their section 5.1), much larger than the escape velocity of the MW near the Solar circle. This is why their Figure 9 demonstrates that EMOND should have only a very small effect on the dynamics of relatively isolated galaxies, preserving the successes of standard MOND in such systems \citep[e.g.][]{Lelli_2017}. Therefore, our calculated escape velocity curve for the MW assuming standard MOND should also be very nearly correct in EMOND.

\subsection{The hot gas corona of the Milky Way}
\subsubsection{Corona mass}
\begin{figure}
	\includegraphics [width = 8.5cm] {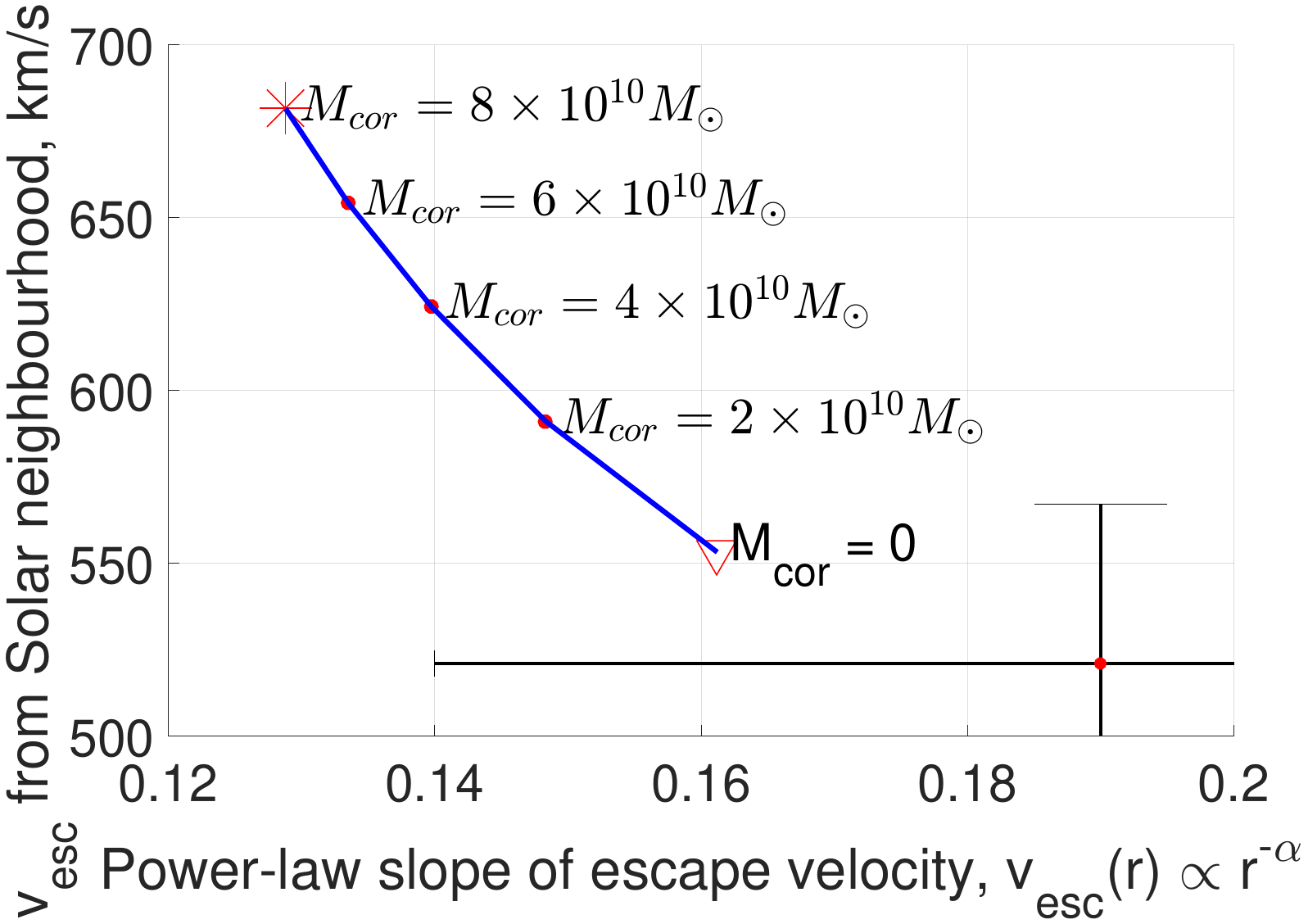}
	\caption{Effect of the MW corona mass $M_{_{cor}}$ on its escape velocity curve for points along its disk axis in the direction of the external field. We use the same model as in Figure \ref{v_esc_profile}. The $x$-axis shows the value of $\alpha$ such that $v_{esc} \left( r \right) ~\propto~ r^{-\alpha}$ while the $y$-axis shows $v_{esc}$ near the Sun. The measured values of these quantities are shown as a red dot with black error bars towards the bottom right \citep{Williams_2017}.}
	\label{v_esc_results_top_axis_no_corona}
\end{figure}

It is clear that our calculated escape velocities are towards the upper end of the range allowed by observations. Thus, our analysis disfavours a hot gas corona. We have included one because XMM-Newton \citep{Jansen_2001} observations at a range of Galactic latitudes indicate that one is present \citep{Nicastro_2016}. The mass in this corona is unclear, but their best-fitting model suggested $2 \times 10^{10} M_\odot$ (see their Table 2 model A) which we therefore use as our lowest value for $M_{_{cor}}$. Similarly to this work, the best fit to their observations was obtained for the lowest mass corona model they tried out, though substantially more massive halos are far from ruled out. A similar analysis by \citet{Miller_2013} suggests that $M_{_{cor}} \approx 4 \times 10^{10} M_\odot$, which would imply a slightly larger and slower-declining $v_{esc}$ but not to such an extent that observations rule it out (Figure \ref{v_esc_results_top_axis_no_corona}).

A hot gas corona would also be expected to cause ram pressure stripping effects on MW satellites containing gas. This may explain the truncation of the LMC gas disk at a much shorter distance than the extent of its stellar disk \citep{Salem_2015}. Those authors used this argument to estimate that $M_{_{cor}} = 2.7 \pm 1.4 \times 10^{10} M_\odot$, consistent with the other estimates.

Although our analysis is consistent with a corona of this mass, it clearly prefers an even lower mass. We therefore investigated the effect of lowering $M_{_{cor}}$ all the way down to 0. As expected, this would make the MW $v_{esc}$ curve slightly more consistent with observations in terms of both its amplitude and its radial gradient (Figure \ref{v_esc_results_top_axis_no_corona}).

The corona of the MW would affect its satellites not only through ram pressure stripping but also by modifying their orbits, especially if the corona mass was significant compared to that of the MW stellar and gas disks. This led \citet{Thomas_2017} to investigate whether a MOND model of the Sagittarius tidal stream \citep{Newberg_2002} was more consistent if a corona is included. The orbit of the progenitor would indeed be rather different with a massive corona, though the mass tried by \citet{Thomas_2017} was rather high $\left( 1.5 \times 10^{11} M_\odot \right)$. This led to tidal stream radial velocities that are more consistent with observations. However, the higher MW mass also led to a much smaller apocentre for the trailing arm, making it very difficult to explain the observed distance to its apocentre \citep{Belokurov_2014}. In future, a critical objective will be to see if the stars corresponding to these measurements are indeed part of the Sagittarius tidal stream, which is sometimes difficult to detect against foreground and background stars. For the time being, almost all observations of it remain consistent with a detailed MOND model of the MW that is not surrounded by a massive corona.

\subsubsection{Corona scale length}
\label{Corona_scale_length}
The scale length of the corona is harder to pin down because spectroscopic observations are usually only sensitive to the total amount of an absorber integrated along a particular line of sight. Nonetheless, the corona can't be too centrally concentrated if it stripped the LMC gas disk \citep{Salem_2015}. This suggests a scale length comparable to the 50 kpc distance of the LMC \citep{Pietrzynski_2013}. Much larger values would cause a substantial part of the corona to lie closer to M31 than to the MW (Equation \ref{Plummer_distribution}). This is unlikely given that the spatial distribution and redshifts of absorbing material in the corona strongly suggests that it surrounds the MW rather than the LG as a whole \citep{Bregman_2007}. This led us to explore values for $r_{_{cor}}$ in the range $\left( 20 - 60 \right)$ kpc. Our results are not much affected by this parameter (Figure \ref{v_esc_results_top_axis}).

\subsection{The external field strength}

The time-integrated effect of an EF on the LG must be that it now has a peculiar velocity $v_{pec}$ with respect to the average matter distribution in the Universe. Assuming the direction of the EF has always remained the same, we must have that
\begin{eqnarray}
	\int_{0}^{t_f} a~g_{ext}~dt ~=~ v_{pec} \, .
	\label{g_ext_estimation}
\end{eqnarray}

Here, $t$ is the time elapsed since the Big Bang, with present value $t_f$. The integrating factor $a$ is required to account for the effect of Hubble drag, which arises because objects tend to move into regions where the Hubble flow velocity is more nearly equal to the velocity of the object \citep[e.g.][equation 24]{Banik_Zhao_2016}. The LG currently has a peculiar velocity of $\ssim 630$ km/s with respect to the surface of last scattering \citep{Kogut_1993}, suggesting that $g_{ext} \approx 0.015 a_{_0}$. However, the present EF on the MW may be different to this as only its time integral is constrained and because we also need to consider other objects in the LG when estimating the EF on the MW as opposed to the EF on the LG as a whole. The most obvious such object is M31, whose sky position (Galactic co-ordinates $121.2^\circ, -21.6^\circ $) is roughly opposite the peculiar velocity of the LG as a whole $\left(276^\circ, 30^\circ \right)$. Dividing the square of the 225 km/s flatline rotation curve amplitude of M31 \citep{Carignan_2006} by its 783 kpc distance \citep{McConnachie_2012} suggests that the resulting EF on the MW nearly cancels that due to objects outside the LG.

However, there could be non-negligible contributions arising from other objects much closer to the MW such as its satellites. The brightest MW satellite is the LMC, only $\ssim 50$ kpc away \citep{Pietrzynski_2013}. To estimate its mass, we note that its rotation curve flatline level is $\ssim 90$ km/s \citep{Kallivayalil_2013}, only $\ssim \frac{1}{2}$ that of the MW \citep{Kafle_2012}. Thus, the LMC is likely only $\ssim \left( \frac{1}{2} \right)^4 = \frac{1}{16}$ as massive as the MW. The gravitational field of the LMC on the MW can be estimated by using Newton's Third Law, which still works in QUMOND as the theory can be derived from a least action principle \citep{QUMOND}. The gravity of the MW at a distance of $r = 50$ kpc is $\sim \frac{v^2}{r}$ with $v \approx 180$ km/s, suggesting that the recoil acceleration of the MW is only $\ssim 0.01 a_{_0}$.

Of course, the LMC is not far enough away that its effect on the MW can be considered as a constant EF. Nonetheless, this gives an idea of how significant the LMC could be for our analysis, at least for $v_{esc}$ in the Solar neighbourhood. Further away, the effect of the LMC could be larger or smaller depending on the direction. As the SDSS images the sky from the northern hemisphere while the LMC has a declination of $-70^\circ$, it should not significantly affect the $v_{esc}$ measurements of \citet{Williams_2017} that we use to constrain our models. Even so, it could lead to a larger EF on the survey volume, perhaps explaining why our models prefer slightly higher EF strengths than suggested by Equation \ref{g_ext_estimation}. Of course, this equation does not constrain $g_{_{ext}}$ all that well because we do not know the time dependence of the EF on the LG. It could well be stronger now than its typical strength over cosmic history due to motion of objects outside the LG e.g. if their gravitational fields on it cancelled out in the past but no longer do so as a result of recent structure formation. Thus, it is quite possible that $g_{_{ext}} \approx {0.03 a_{_{0}}}$, as suggested by our analysis.

The high-velocity stars relevant to a $v_{esc}$ measurement are expected to have a rather long orbital time and thus `remember' conditions several Gyr ago. At that time, the smaller scale-factor of the Universe implies that large-scale structures were closer to the MW. One expects the structures themselves to be less pronounced at earlier times, but structure formation has been slowed down by the effect of dark energy and mostly happened at much earlier cosmological epochs. Thus, $g_{_{ext}}$ would likely have been larger a few Gyr ago, leading to a shallower potential well around the MW. Although this would have subsequently deepened, it may have done so too quickly for the velocity function of long-period MW stars to expand into the newly allowed region. Consequently, the observationally determined $v_{esc}$ might well fall short of its true value.

%
%

\subsection{The external field direction}
\label{External_field_direction}

To estimate how $\bm{\widehat{g}_{_{ext}}}$ affects our results, we note that the EF only substantially affects the potential rather far from the MW (Equation \ref{r_ext}). At such large distances, it becomes appropriate to consider it as a point mass. This allows us to estimate how much our escape velocity calculations could be affected by $\bm{\widehat{g}_{_{ext}}}$.

We begin with a simple analytic demonstration that the effect is not likely to be large. To do this, we consider the potential difference between equally distant points located along $\bm{\widehat{g}_{_{ext}}}$ and at right angles. The distance we consider is $r_{_{ext}}$ (Equation \ref{r_ext}) because, at closer distances, the EF is not dominant. Thus, for $r \la r_{_{ext}}$, we essentially have an isolated point mass, whose gravitational field is of course spherically symmetric and unaffected by $\bm{\widehat{g}_{_{ext}}}$. We use the analytic results of \citet{Banik_Zhao_2015} to obtain that
\begin{eqnarray}
	\left.\Delta \Phi \right|_{\theta = 0, \frac{\pi}{2}} = \frac{GM\nu_{_{ext}}}{4r_{_{ext}}} ~~\text{ at } r = r_{_{ext}} \, .
	\label{Delta_Phi_r_ext}
\end{eqnarray}

This is the amount by which the potential is expected to be deeper along $\pm \bm{\widehat{g}_{_{ext}}}$ than at right angles. However, it only applies at a distance of $r = r_{_{ext}}$ whereas we are interested in $\left.\Delta \Phi \right|_{\theta = 0, \frac{\pi}{2}}$ at $r = R_\odot$. If we assume that the tangential gravity $\bm{g}_{_\theta}$ has a similar magnitude throughout the nearly isolated region $r \la r_{_{ext}}$, then this implies that 
\begin{eqnarray}
	\left.\Delta \Phi \right|_{\theta = 0, \frac{\pi}{2}} = \frac{GM\nu_{_{ext}}R_\odot}{4{r_{_{ext}}}^2}~~\text{ at } r = R_\odot \, .
\end{eqnarray}

For $v_{esc} \sim 600$ km/s, this yields an effect of $\ssim 400$ m/s. In reality, $\bm{g}_{_\theta}$ could be substantially smaller than we assumed. However, it could not be much larger because we expect $\bm{g}$ to be almost spherically symmetric near a point mass. As the radial component of $\bm{g} \appropto r^{-1}$ for $r \la r_{_{ext}}$\footnote{the DML is a reasonable assumption beyond the Solar circle}, its tangential component must increase inwards at a much lower rate, if at all. It can of course remain flat as just assumed or actually decrease inwards e.g. $\bm{g}_{_\theta} \propto r^1$. This yields an uncertainty of $\ssim 40 \times$ to our estimate of how much $v_{esc}$ varies with $\bm{\widehat{g}_{_{ext}}}$ because $R_\odot \approx 0.025 r_{_{ext}}$.

Even so, it is clear that the effect of $\bm{\widehat{g}_{_{ext}}}$ does not exceed 20 km/s, a very conservative upper limit that would be correct only if $\bm{g}_{_\theta} \propto r^{-1}$ for $r < r_{_{ext}}$. In this case, Equation \ref{Delta_Phi_r_ext} would remain valid at the Solar circle, implying a gravitational field that deviates from spherical symmetry by $\ssim \frac{1}{4}$ at $r = R_\odot$ even though this is only a few percent of $r_{_{ext}}$. At this position, it is highly unlikely for the gravity of a point mass to deviate this far from spherical symmetry. Nonetheless, it corresponds to a previous estimate of how much $\bm{\widehat{g}_{_{ext}}}$ might influence $v_{esc}$ near the Sun \citep[][section 3.2]{Famaey_2007}.

\begin{figure}
	\centering 
	\includegraphics [width = 8.5cm] {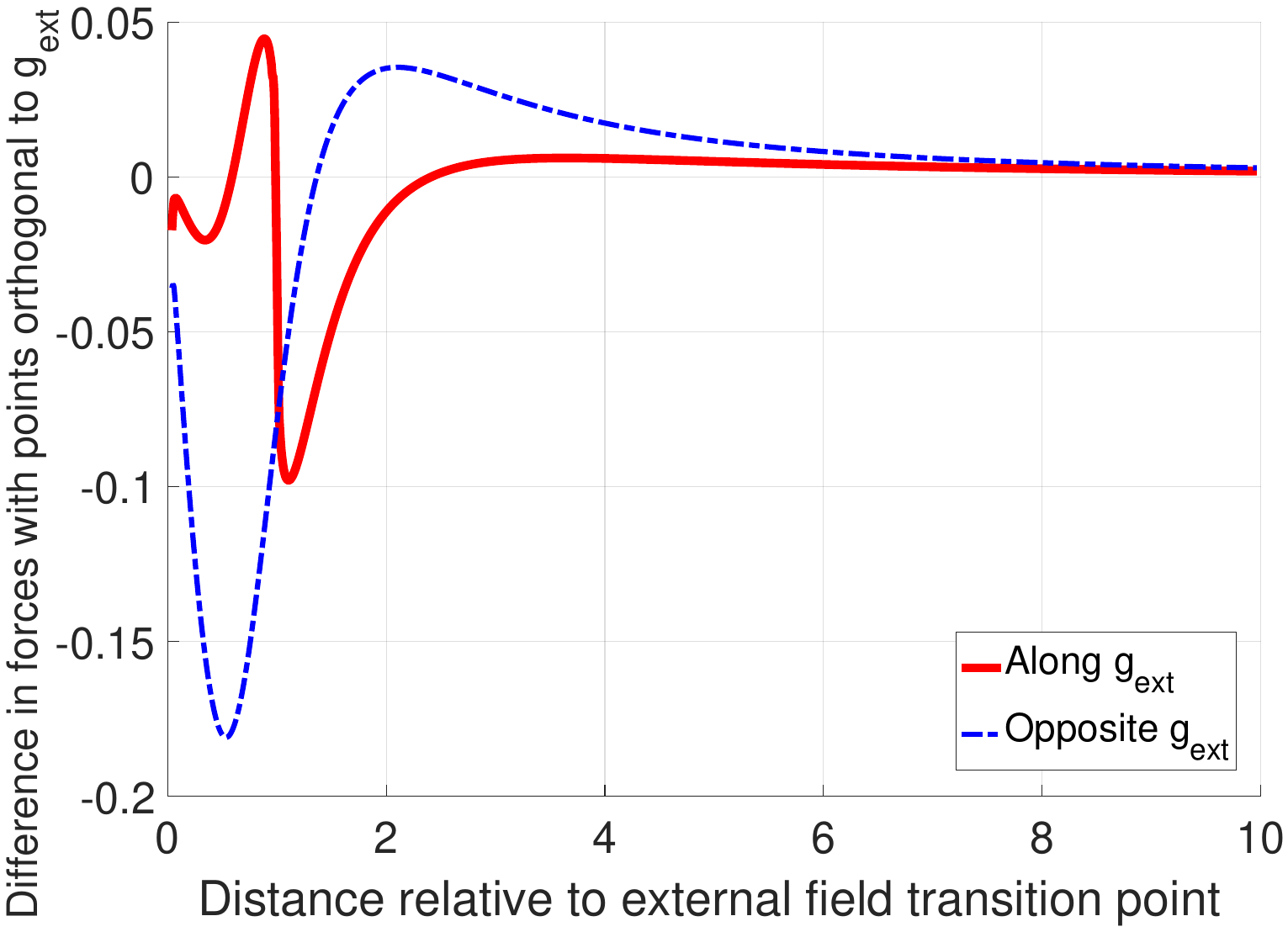}
	\caption{The difference between the radial component of the gravitational field in the directions along and orthogonal to the external field, for a point mass in the deep-MOND limit. The units are such that $G = M = a_{_0} = 1$. We show results for points in the direction of $\bm{\widehat{g}_{_{ext}}}$ (solid red) and $-\bm{\widehat{g}_{_{ext}}}$ (dashed blue). In both cases, the force is stronger than in the orthogonal direction at long range, as required by analytic calculations \citep{Banik_Zhao_2015}. However, this is not true at all radii. Notice how the radial forces at short range become very nearly the same in all three directions shown here, even though the gravitational field diverges as $\frac{1}{r}$. This is because the EF is unimportant at $r \ll 1$, reducing the situation to the spherically symmetric case of an isolated point mass.}
	\label{Force_differences_point_g_ext}
\end{figure}

To obtain a better estimate, we look at the radial gravity generated by our point mass $+ g_{_{ext}}$ library and integrate this to get the potential. In this case, we do not need to use a radius of $r_{_{ext}}$ but try to use $R_\odot$ instead. Unfortunately, this is a very small fraction of $r_{_{ext}}$ and our algorithm is unable to resolve such a small scale. Thus, we use a radius of 0.04$r_{_{ext}}$, which we consider acceptable because the EF is $>25 \times$ weaker than the gravity of the point mass in such regions. In reality, it is the Newtonian gravity $\bm{g_{_N}}$ rather than the actual gravity $\bm{g}$ which determines the phantom dark matter density that must be integrated to get $\bm{g}$ (Equation \ref{QUMOND_governing_equation}). At these distances, $\bm{g_{_N}}$ from the point mass would be $> 625 \times$ stronger than the EF, implying that $\bm{g}$ should be almost spherically symmetric. Moreover, the missing radial range is rather small and thus unlikely to cause a large error in the depth of the potential.

In this way, we find that $\left.\Delta \Phi \right|_{\theta = 0, \frac{\pi}{2}}$ corresponds to a velocity of $\ssim 3$ km/s which would only negligibly affect our escape velocity calculations ($\Delta v_{esc} \sim 10$ m/s). To see why the effect is so small, we look at the radial component of the gravity generated by a point mass in the directions parallel and orthogonal to $\bm{\widehat{g}_{_{ext}}}$. Although the forces are stronger along $\bm{\widehat{g}_{_{ext}}}$ at large radii, they are \emph{weaker} at small radii, leading to partial cancellation (Figure \ref{Force_differences_point_g_ext}). This is true both for points along $\bm{\widehat{g}_{_{ext}}}$ and in the opposite direction, albeit in slightly different ways. Therefore, we conclude that $\bm{\widehat{g}_{_{ext}}}$ should not much affect our results, making it sufficient to consider the computationally much simpler case where $\bm{\widehat{g}_{_{ext}}}$ is aligned with the symmetry axis of the MW disk.

Our calculations show that the gravitational field due to a point mass is not symmetric between polar angles of $\theta$ and $\pi - \theta$, even if the potential very nearly is. There may be other situations where this asymmetry does have an effect. For example, a tidal stream due to a low mass MW satellite could be considered as embedded in the EF of the MW but also feeling the gravity of the progenitor. If the two are comparable, then it is likely that the tidal stream would develop asymmetrically due to the asymmetric gravitational field. For an individual object, the asymmetry could perhaps be explained in other ways such as a passing dark matter subhalo. Nonetheless, the EFE must always work in a certain way in MOND whereas a subhalo could pass by the progenitor on either side. Thus, the observation of a large number of such tidal streams could help to distinguish between the models. Unfortunately, the Sagittarius tidal stream does not seem to be much affected by the EFE in fully consistent MOND simulations of it \citep{Thomas_2017}, but the theory may predict distinctive behaviour due to the EFE in other tidal streams.

\section{Conclusions}
\label{Conclusions}

In the context of Modified Newtonian Dynamics (MOND), we investigated the escape velocity curve of the Milky Way (MW) using a wide range of baryonic models (Table \ref{Parameters}). A reasonably good fit is found over the observed region \citep[Galactocentric radii 8$-$50 kpc,][]{Williams_2017} using a plausible model where the MW stellar and gas disks have scale lengths fixed by observations and masses consistent with the observed MW rotation curve (Figures \ref{v_esc_disk_results} and \ref{v_esc_results_top_axis}). The required external field strength is $\ssim 0.03 a_{_0}$ or slightly higher while a fairly low mass corona is preferred (Figure \ref{v_esc_results_top_axis_no_corona}). Indeed, our best fits are obtained if the corona does not exist at all (Figure \ref{v_esc_results_top_axis_no_corona}). As this is unlikely on other grounds \citep[e.g.][]{Nicastro_2016}, it is important to note that our analysis is consistent with a corona mass of $2 \times 10^{10} M_\odot$, near the lower limit of the range allowed by independent observations (see their Table 2 model A). Our results should not be much affected by the spatial extent of the MW corona (Section \ref{Corona_scale_length}) or the direction of the external field on it (Section \ref{External_field_direction}).

Using just the directly observed baryonic mass of the MW, it is possible to understand both the radial gradient of its potential and its absolute depth, as measured by its circular and escape velocity curves, respectively. It will be exciting to see whether such a model remains consistent with observations in the GAIA era.


\section{Acknowledgements}
\label{Acknowledgements}

IB is supported by Science and Technology Facilities Council studentship 1506672. He is grateful to Stacy McGaugh for suggesting appropriate mass models for the Milky Way.
The algorithms were set up using \textsc{matlab}$^\text{\textregistered}$.

\bibliographystyle{mnras}
\bibliography{EVM_bbl}

\begin{thebibliography}{}
\makeatletter
\relax
\def\mn@urlcharsother{\let\do\@makeother \do\$\do\&\do\#\do\^\do\_\do\%\do\~}
\def\mn@doi{\begingroup\mn@urlcharsother \@ifnextchar [ {\mn@doi@}
  {\mn@doi@[]}}
\def\mn@doi@[#1]#2{\def\@tempa{#1}\ifx\@tempa\@empty \href
  {http://dx.doi.org/#2} {doi:#2}\else \href {http://dx.doi.org/#2} {#1}\fi
  \endgroup}
\def\mn@eprint#1#2{\mn@eprint@#1:#2::\@nil}
\def\mn@eprint@arXiv#1{\href {http://arxiv.org/abs/#1} {{\tt arXiv:#1}}}
\def\mn@eprint@dblp#1{\href {http://dblp.uni-trier.de/rec/bibtex/#1.xml}
  {dblp:#1}}
\def\mn@eprint@#1:#2:#3:#4\@nil{\def\@tempa {#1}\def\@tempb {#2}\def\@tempc
  {#3}\ifx \@tempc \@empty \let \@tempc \@tempb \let \@tempb \@tempa \fi \ifx
  \@tempb \@empty \def\@tempb {arXiv}\fi \@ifundefined
  {mn@eprint@\@tempb}{\@tempb:\@tempc}{\expandafter \expandafter \csname
  mn@eprint@\@tempb\endcsname \expandafter{\@tempc}}}

\bibitem[\protect\citeauthoryear{{Ahn} et~al.,}{{Ahn} et~al.}{2012}]{Ahn_2012}
{Ahn} C.~P.,  et~al., 2012, \mn@doi [ApJS] {10.1088/0067-0049/203/2/21}, \href
  {http://adsabs.harvard.edu/abs/2012ApJS..203...21A} {203, 21}

\bibitem[\protect\citeauthoryear{{Banik}}{{Banik}}{2014}]{Banik_2014}
{Banik} I.,  2014, preprint, \href
  {http://adsabs.harvard.edu/abs/2014arXiv1406.4538B} {Arxiv} (\mn@eprint
  {arXiv} {1406.4538v2})

\bibitem[\protect\citeauthoryear{{Banik} \& {Zhao}}{{Banik} \&
  {Zhao}}{2016}]{Banik_Zhao_2016}
{Banik} I.,  {Zhao} H.,  2016, \mn@doi [MNRAS] {10.1093/mnras/stw787}, \href
  {http://adsabs.harvard.edu/abs/2016MNRAS.459.2237B} {459, 2237}

\bibitem[\protect\citeauthoryear{{Banik} \& {Zhao}}{{Banik} \&
  {Zhao}}{2017a}]{Banik_Zhao_2017}
{Banik} I.,  {Zhao} H.,  2017a, \mn@doi [MNRAS] {10.1093/mnras/stx151}, \href
  {http://adsabs.harvard.edu/abs/2017MNRAS.tmp..165B} {467, 2180}

\bibitem[\protect\citeauthoryear{{Banik} \& {Zhao}}{{Banik} \&
  {Zhao}}{2017b}]{Banik_2017_anisotropy}
{Banik} I.,  {Zhao} H.,  2017b, \mn@doi [MNRAS] {10.1093/mnras/stx2596}, \href
  {http://adsabs.harvard.edu/abs/2017arXiv170106559B} {stx2596}

\bibitem[\protect\citeauthoryear{{Banik} \& {Zhao}}{{Banik} \&
  {Zhao}}{2018}]{Banik_Zhao_2015}
{Banik} I.,  {Zhao} H.,  2018, SciFed Journal of Astrophysics, \href
  {https://ui.adsabs.harvard.edu/abs/2015arXiv150908457B} {1, 1000008}

\bibitem[\protect\citeauthoryear{{Bekenstein} \& {Milgrom}}{{Bekenstein} \&
  {Milgrom}}{1984}]{Bekenstein_Milgrom_1984}
{Bekenstein} J.,  {Milgrom} M.,  1984, \mn@doi [ApJ] {10.1086/162570}, \href
  {http://adsabs.harvard.edu/abs/1984ApJ...286....7B} {286, 7}

\bibitem[\protect\citeauthoryear{{Bell} \& {de Jong}}{{Bell} \& {de
  Jong}}{2001}]{Bell_de_Jong_2001}
{Bell} E.~F.,  {de Jong} R.~S.,  2001, \mn@doi [ApJ] {10.1086/319728}, \href
  {http://adsabs.harvard.edu/abs/2001ApJ...550..212B} {550, 212}

\bibitem[\protect\citeauthoryear{{Belokurov} et~al.,}{{Belokurov}
  et~al.}{2014}]{Belokurov_2014}
{Belokurov} V.,  et~al., 2014, \mn@doi [MNRAS] {10.1093/mnras/stt1862}, \href
  {http://adsabs.harvard.edu/abs/2014MNRAS.437..116B} {437, 116}

\bibitem[\protect\citeauthoryear{{Bovy} \& {Rix}}{{Bovy} \&
  {Rix}}{2013}]{Bovy_2013}
{Bovy} J.,  {Rix} H.-W.,  2013, \mn@doi [ApJ] {10.1088/0004-637X/779/2/115},
  \href {http://adsabs.harvard.edu/abs/2013ApJ...779..115B} {779, 115}

\bibitem[\protect\citeauthoryear{{Brada} \& {Milgrom}}{{Brada} \&
  {Milgrom}}{1999}]{Brada_1999}
{Brada} R.,  {Milgrom} M.,  1999, \mn@doi [ApJ] {10.1086/307402}, \href
  {http://adsabs.harvard.edu/abs/1999ApJ...519..590B} {519, 590}

\bibitem[\protect\citeauthoryear{{Bregman} \& {Lloyd-Davies}}{{Bregman} \&
  {Lloyd-Davies}}{2007}]{Bregman_2007}
{Bregman} J.~N.,  {Lloyd-Davies} E.~J.,  2007, \mn@doi [ApJ] {10.1086/521321},
  \href {http://adsabs.harvard.edu/abs/2007ApJ...669..990B} {669, 990}

\bibitem[\protect\citeauthoryear{{Brunthaler}, {Reid}, {Falcke}, {Henkel}  \&
  {Menten}}{{Brunthaler} et~al.}{2007}]{Brunthaler_2007}
{Brunthaler} A.,  {Reid} M.~J.,  {Falcke} H.,  {Henkel} C.,   {Menten} K.~M.,
  2007, \mn@doi [A\&A] {10.1051/0004-6361:20066430}, \href
  {http://adsabs.harvard.edu/abs/2007A\%26A...462..101B} {462, 101}

\bibitem[\protect\citeauthoryear{{Caldwell} et~al.,}{{Caldwell}
  et~al.}{2017}]{Caldwell_2017}
{Caldwell} N.,  et~al., 2017, \mn@doi [ApJ] {10.3847/1538-4357/aa688e}, \href
  {http://adsabs.harvard.edu/abs/2017ApJ...839...20C} {839, 20}

\bibitem[\protect\citeauthoryear{{Candlish}, {Smith}  \&
  {Fellhauer}}{{Candlish} et~al.}{2015}]{Candlish_2015}
{Candlish} G.~N.,  {Smith} R.,   {Fellhauer} M.,  2015, \mn@doi [MNRAS]
  {10.1093/mnras/stu2158}, \href
  {http://adsabs.harvard.edu/abs/2015MNRAS.446.1060C} {446, 1060}

\bibitem[\protect\citeauthoryear{{Carignan}, {Chemin}, {Huchtmeier}  \&
  {Lockman}}{{Carignan} et~al.}{2006}]{Carignan_2006}
{Carignan} C.,  {Chemin} L.,  {Huchtmeier} W.~K.,   {Lockman} F.~J.,  2006,
  \mn@doi [ApJL] {10.1086/503869}, \href
  {http://adsabs.harvard.edu/abs/2006ApJ...641L.109C} {641, L109}

\bibitem[\protect\citeauthoryear{{Chatzopoulos}, {Fritz}, {Gerhard},
  {Gillessen}, {Wegg}, {Genzel}  \& {Pfuhl}}{{Chatzopoulos}
  et~al.}{2015}]{Chatzopoulos_2015}
{Chatzopoulos} S.,  {Fritz} T.~K.,  {Gerhard} O.,  {Gillessen} S.,  {Wegg} C.,
  {Genzel} R.,   {Pfuhl} O.,  2015, \mn@doi [MNRAS] {10.1093/mnras/stu2452},
  \href {http://adsabs.harvard.edu/abs/2015MNRAS.447..948C} {447, 948}

\bibitem[\protect\citeauthoryear{{Famaey} \& {Binney}}{{Famaey} \&
  {Binney}}{2005}]{Famaey_Binney_2005}
{Famaey} B.,  {Binney} J.,  2005, \mn@doi [MNRAS]
  {10.1111/j.1365-2966.2005.09474.x}, \href
  {http://adsabs.harvard.edu/abs/2005MNRAS.363..603F} {363, 603}

\bibitem[\protect\citeauthoryear{{Famaey} \& {McGaugh}}{{Famaey} \&
  {McGaugh}}{2012}]{Famaey_McGaugh_2012}
{Famaey} B.,  {McGaugh} S.~S.,  2012, \mn@doi [Living Reviews in Relativity]
  {10.12942/lrr-2012-10}, \href
  {http://adsabs.harvard.edu/abs/2012LRR....15...10F} {15, 10}

\bibitem[\protect\citeauthoryear{{Famaey}, {Bruneton}  \& {Zhao}}{{Famaey}
  et~al.}{2007}]{Famaey_2007}
{Famaey} B.,  {Bruneton} J.-P.,   {Zhao} H.,  2007, \mn@doi [MNRAS]
  {10.1111/j.1745-3933.2007.00308.x}, \href
  {http://adsabs.harvard.edu/abs/2007MNRAS.377L..79F} {377, L79}

\bibitem[\protect\citeauthoryear{{Gilmore} \& {Reid}}{{Gilmore} \&
  {Reid}}{1983}]{Gilmore_1983}
{Gilmore} G.,  {Reid} N.,  1983, \mn@doi [MNRAS] {10.1093/mnras/202.4.1025},
  \href {http://adsabs.harvard.edu/abs/1983MNRAS.202.1025G} {202, 1025}

\bibitem[\protect\citeauthoryear{{Haghi}, {Bazkiaei}, {Zonoozi}  \&
  {Kroupa}}{{Haghi} et~al.}{2016}]{Haghi_2016}
{Haghi} H.,  {Bazkiaei} A.~E.,  {Zonoozi} A.~H.,   {Kroupa} P.,  2016, \mn@doi
  [MNRAS] {10.1093/mnras/stw573}, \href
  {http://adsabs.harvard.edu/abs/2016MNRAS.458.4172H} {458, 4172}

\bibitem[\protect\citeauthoryear{{Hayden} et~al.,}{{Hayden}
  et~al.}{2015}]{Hayden_2015}
{Hayden} M.~R.,  et~al., 2015, \mn@doi [ApJ] {10.1088/0004-637X/808/2/132},
  \href {http://adsabs.harvard.edu/abs/2015ApJ...808..132H} {808, 132}

\bibitem[\protect\citeauthoryear{{Hodson} \& {Zhao}}{{Hodson} \&
  {Zhao}}{2017}]{Hodson_2017}
{Hodson} A.~O.,  {Zhao} H.,  2017, \mn@doi [A\&A]
  {10.1051/0004-6361/201629358}, \href
  {http://adsabs.harvard.edu/abs/2017A%26A...598A.127H} {598, A127}

\bibitem[\protect\citeauthoryear{{Jansen} et~al.,}{{Jansen}
  et~al.}{2001}]{Jansen_2001}
{Jansen} F.,  et~al., 2001, \mn@doi [A\&A] {10.1051/0004-6361:20000036}, \href
  {http://adsabs.harvard.edu/abs/2001A%26A...365L...1J} {365, L1}

\bibitem[\protect\citeauthoryear{{Jayaraman}, {Gilmore}, {Wyse}, {Norris}  \&
  {Belokurov}}{{Jayaraman} et~al.}{2013}]{Jayaraman_2013}
{Jayaraman} A.,  {Gilmore} G.,  {Wyse} R.~F.~G.,  {Norris} J.~E.,   {Belokurov}
  V.,  2013, \mn@doi [MNRAS] {10.1093/mnras/stt221}, \href
  {http://adsabs.harvard.edu/abs/2013MNRAS.431..930J} {431, 930}

\bibitem[\protect\citeauthoryear{{Juri{\'c}} et~al.,}{{Juri{\'c}}
  et~al.}{2008}]{Juric_2008}
{Juri{\'c}} M.,  et~al., 2008, \mn@doi [ApJ] {10.1086/523619}, \href
  {http://adsabs.harvard.edu/abs/2008ApJ...673..864J} {673, 864}

\bibitem[\protect\citeauthoryear{{Kafle}, {Sharma}, {Lewis}  \&
  {Bland-Hawthorn}}{{Kafle} et~al.}{2012}]{Kafle_2012}
{Kafle} P.~R.,  {Sharma} S.,  {Lewis} G.~F.,   {Bland-Hawthorn} J.,  2012,
  \mn@doi [ApJ] {10.1088/0004-637X/761/2/98}, \href
  {http://adsabs.harvard.edu/abs/2012ApJ...761...98K} {761, 98}

\bibitem[\protect\citeauthoryear{{Kallivayalil}, {van der Marel}, {Besla},
  {Anderson}  \& {Alcock}}{{Kallivayalil} et~al.}{2013}]{Kallivayalil_2013}
{Kallivayalil} N.,  {van der Marel} R.~P.,  {Besla} G.,  {Anderson} J.,
  {Alcock} C.,  2013, \mn@doi [ApJ] {10.1088/0004-637X/764/2/161}, \href
  {http://adsabs.harvard.edu/abs/2013ApJ...764..161K} {764, 161}

\bibitem[\protect\citeauthoryear{{Kogut} et~al.,}{{Kogut}
  et~al.}{1993}]{Kogut_1993}
{Kogut} A.,  et~al., 1993, \mn@doi [ApJ] {10.1086/173453}, \href
  {http://cdsads.u-strasbg.fr/abs/1993ApJ...419....1K} {419, 1}

\bibitem[\protect\citeauthoryear{{Lehner}, {Howk}  \& {Wakker}}{{Lehner}
  et~al.}{2015}]{Lehner_2015}
{Lehner} N.,  {Howk} J.~C.,   {Wakker} B.~P.,  2015, \mn@doi [ApJ]
  {10.1088/0004-637X/804/2/79}, \href
  {http://adsabs.harvard.edu/abs/2015ApJ...804...79L} {804, 79}

\bibitem[\protect\citeauthoryear{{Lelli}, {McGaugh}, {Schombert}  \&
  {Pawlowski}}{{Lelli} et~al.}{2017}]{Lelli_2017}
{Lelli} F.,  {McGaugh} S.~S.,  {Schombert} J.~M.,   {Pawlowski} M.~S.,  2017,
  \mn@doi [ApJ] {10.3847/1538-4357/836/2/152}, \href
  {http://adsabs.harvard.edu/abs/2017ApJ...836..152L} {836, 152}

\bibitem[\protect\citeauthoryear{{Leonard} \& {Tremaine}}{{Leonard} \&
  {Tremaine}}{1990}]{Leonard_1990}
{Leonard} P.~J.~T.,  {Tremaine} S.,  1990, \mn@doi [ApJ] {10.1086/168638},
  \href {http://adsabs.harvard.edu/abs/1990ApJ...353..486L} {353, 486}

\bibitem[\protect\citeauthoryear{{L{\"u}ghausen}, {Famaey}  \&
  {Kroupa}}{{L{\"u}ghausen} et~al.}{2015}]{PoR}
{L{\"u}ghausen} F.,  {Famaey} B.,   {Kroupa} P.,  2015, \mn@doi [Canadian
  Journal of Physics] {10.1139/cjp-2014-0168}, \href
  {http://adsabs.harvard.edu/abs/2015CaJPh..93..232L} {93, 232}

\bibitem[\protect\citeauthoryear{{McConnachie}}{{McConnachie}}{2012}]{McConnachie_2012}
{McConnachie} A.~W.,  2012, \mn@doi [AJ] {10.1088/0004-6256/144/1/4}, \href
  {http://adsabs.harvard.edu/abs/2012AJ....144....4M} {144, 4}

\bibitem[\protect\citeauthoryear{{McGaugh}}{{McGaugh}}{2008}]{McGaugh_2008}
{McGaugh} S.~S.,  2008, \mn@doi [ApJ] {10.1086/589148}, \href
  {http://adsabs.harvard.edu/abs/2008ApJ...683..137M} {683, 137}

\bibitem[\protect\citeauthoryear{{McGaugh}}{{McGaugh}}{2011}]{McGaugh_2011}
{McGaugh} S.~S.,  2011, \mn@doi [Physical Review Letters]
  {10.1103/PhysRevLett.106.121303}, \href
  {http://adsabs.harvard.edu/abs/2011PhRvL.106l1303M} {106, 121303}

\bibitem[\protect\citeauthoryear{{McGaugh}}{{McGaugh}}{2016a}]{McGaugh_2016_MW}
{McGaugh} S.~S.,  2016a, \mn@doi [ApJ] {10.3847/0004-637X/816/1/42}, \href
  {http://adsabs.harvard.edu/abs/2016ApJ...816...42M} {816, 42}

\bibitem[\protect\citeauthoryear{{McGaugh}}{{McGaugh}}{2016b}]{McGaugh_2016_Crater}
{McGaugh} S.~S.,  2016b, \mn@doi [ApJL] {10.3847/2041-8205/832/1/L8}, \href
  {http://adsabs.harvard.edu/abs/2016ApJ...832L...8M} {832, L8}

\bibitem[\protect\citeauthoryear{{McGaugh}, {Lelli}  \& {Schombert}}{{McGaugh}
  et~al.}{2016}]{McGaugh_Lelli_2016}
{McGaugh} S.,  {Lelli} F.,   {Schombert} J.,  2016, \mn@doi [Phys. Rev. Lett.]
  {10.1103/PhysRevLett.117.201101}, \href
  {http://adsabs.harvard.edu/abs/2016arXiv160905917M} {117, 201101}

\bibitem[\protect\citeauthoryear{{McMillan}}{{McMillan}}{2017}]{McMillan_2017}
{McMillan} P.~J.,  2017, \mn@doi [MNRAS] {10.1093/mnras/stw2759}, \href
  {http://adsabs.harvard.edu/abs/2017MNRAS.465...76M} {465, 76}

\bibitem[\protect\citeauthoryear{{Meillon}, {Crifo}, {Gomez}, {Udry}  \&
  {Mayor}}{{Meillon} et~al.}{1998}]{Meillon_1998}
{Meillon} L.,  {Crifo} F.,  {Gomez} A.~E.,  {Udry} S.,   {Mayor} M.,  1998, in
  {Zaritsky} D.,  ed.,  Astronomical Society of the Pacific Conference Series
  Vol. 136, Galactic Halos. p.~230

\bibitem[\protect\citeauthoryear{{Milgrom}}{{Milgrom}}{1983}]{Milgrom_1983}
{Milgrom} M.,  1983, \mn@doi [ApJ] {10.1086/161130}, \href
  {http://adsabs.harvard.edu/abs/1983ApJ...270..365M} {270, 365}

\bibitem[\protect\citeauthoryear{{Milgrom}}{{Milgrom}}{1986}]{Milgrom_1986}
{Milgrom} M.,  1986, \mn@doi [ApJ] {10.1086/164021}, \href
  {http://adsabs.harvard.edu/abs/1986ApJ...302..617M} {302, 617}

\bibitem[\protect\citeauthoryear{{Milgrom}}{{Milgrom}}{1999}]{Milgrom_1999}
{Milgrom} M.,  1999, \mn@doi [Phys. Lett. A] {10.1016/S0375-9601(99)00077-8},
  \href {http://adsabs.harvard.edu/abs/1999PhLA..253..273M} {253, 273}

\bibitem[\protect\citeauthoryear{{Milgrom}}{{Milgrom}}{2010}]{QUMOND}
{Milgrom} M.,  2010, \mn@doi [MNRAS] {10.1111/j.1365-2966.2009.16184.x}, \href
  {http://adsabs.harvard.edu/abs/2010MNRAS.403..886M} {403, 886}

\bibitem[\protect\citeauthoryear{{Miller} \& {Bregman}}{{Miller} \&
  {Bregman}}{2013}]{Miller_2013}
{Miller} M.~J.,  {Bregman} J.~N.,  2013, \mn@doi [ApJ]
  {10.1088/0004-637X/770/2/118}, \href
  {http://adsabs.harvard.edu/abs/2013ApJ...770..118M} {770, 118}

\bibitem[\protect\citeauthoryear{{Newberg} et~al.,}{{Newberg}
  et~al.}{2002}]{Newberg_2002}
{Newberg} H.~J.,  et~al., 2002, \mn@doi [ApJ] {10.1086/338983}, \href
  {http://adsabs.harvard.edu/abs/2002ApJ...569..245N} {569, 245}

\bibitem[\protect\citeauthoryear{{Nicastro}, {Senatore}, {Krongold}, {Mathur}
  \& {Elvis}}{{Nicastro} et~al.}{2016}]{Nicastro_2016}
{Nicastro} F.,  {Senatore} F.,  {Krongold} Y.,  {Mathur} S.,   {Elvis} M.,
  2016, \mn@doi [ApJL] {10.3847/2041-8205/828/1/L12}, \href
  {http://adsabs.harvard.edu/abs/2016ApJ...828L..12N} {828, L12}

\bibitem[\protect\citeauthoryear{{Norris} et~al.,}{{Norris}
  et~al.}{2016}]{Norris_2016}
{Norris} M.~A.,  et~al., 2016, \mn@doi [\apj] {10.3847/0004-637X/832/2/198},
  \href {http://adsabs.harvard.edu/abs/2016ApJ...832..198N} {832, 198}

\bibitem[\protect\citeauthoryear{{Olling} \& {Merrifield}}{{Olling} \&
  {Merrifield}}{2001}]{Olling_Merrifield_2001}
{Olling} R.~P.,  {Merrifield} M.~R.,  2001, \mn@doi [MNRAS]
  {10.1046/j.1365-8711.2001.04581.x}, \href
  {http://adsabs.harvard.edu/abs/2001MNRAS.326..164O} {326, 164}

\bibitem[\protect\citeauthoryear{{Ostriker} \& {Steinhardt}}{{Ostriker} \&
  {Steinhardt}}{1995}]{Ostriker_1995}
{Ostriker} J.~P.,  {Steinhardt} P.~J.,  1995, \mn@doi [Nature]
  {10.1038/377600a0}, \href {http://adsabs.harvard.edu/abs/1995Natur.377..600O}
  {377, 600}

\bibitem[\protect\citeauthoryear{{Pazy}}{{Pazy}}{2013}]{Pazy_2013}
{Pazy} E.,  2013, \mn@doi [Phys. Rev. D] {10.1103/PhysRevD.87.084063}, \href
  {http://adsabs.harvard.edu/abs/2013PhRvD..87h4063P} {87, 084063}

\bibitem[\protect\citeauthoryear{{Pe{\~n}arrubia}, {Ma}, {Walker}  \&
  {McConnachie}}{{Pe{\~n}arrubia} et~al.}{2014}]{Jorge_2014}
{Pe{\~n}arrubia} J.,  {Ma} Y.-Z.,  {Walker} M.~G.,   {McConnachie} A.,  2014,
  \mn@doi [MNRAS] {10.1093/mnras/stu879}, \href
  {http://adsabs.harvard.edu/abs/2014MNRAS.443.2204P} {443, 2204}

\bibitem[\protect\citeauthoryear{{Perryman}}{{Perryman}}{1989}]{Perryman_1989}
{Perryman} M.~A.~C.,  1989, \mn@doi [Nature] {10.1038/340111a0}, \href
  {http://adsabs.harvard.edu/abs/1989Natur.340..111P} {340, 111}

\bibitem[\protect\citeauthoryear{{Pietrzy{\'n}ski} et~al.,}{{Pietrzy{\'n}ski}
  et~al.}{2013}]{Pietrzynski_2013}
{Pietrzy{\'n}ski} G.,  et~al., 2013, \mn@doi [Nature] {10.1038/nature11878},
  \href {http://adsabs.harvard.edu/abs/2013Natur.495...76P} {495, 76}

\bibitem[\protect\citeauthoryear{{Piffl} et~al.,}{{Piffl}
  et~al.}{2014}]{Piffl_2014}
{Piffl} T.,  et~al., 2014, \mn@doi [A\&A] {10.1051/0004-6361/201322531}, \href
  {http://adsabs.harvard.edu/abs/2014A\%26A...562A..91P} {562, A91}

\bibitem[\protect\citeauthoryear{{Plummer}}{{Plummer}}{1911}]{Plummer_1911}
{Plummer} H.~C.,  1911, \mn@doi [MNRAS] {10.1093/mnras/71.5.460}, \href
  {http://adsabs.harvard.edu/abs/1911MNRAS..71..460P} {71, 460}

\bibitem[\protect\citeauthoryear{{Privon}, {Barnes}, {Evans}, {Hibbard}, {Yun},
  {Mazzarella}, {Armus}  \& {Surace}}{{Privon} et~al.}{2013}]{Privon_2013}
{Privon} G.~C.,  {Barnes} J.~E.,  {Evans} A.~S.,  {Hibbard} J.~E.,  {Yun}
  M.~S.,  {Mazzarella} J.~M.,  {Armus} L.,   {Surace} J.,  2013, \mn@doi [\apj]
  {10.1088/0004-637X/771/2/120}, \href
  {http://adsabs.harvard.edu/abs/2013ApJ...771..120P} {771, 120}

\bibitem[\protect\citeauthoryear{{Quillen} \& {Garnett}}{{Quillen} \&
  {Garnett}}{2001}]{Quillen_2001}
{Quillen} A.~C.,  {Garnett} D.~R.,  2001, in {Funes} J.~G.,  {Corsini} E.~M.,
  eds,  Astronomical Society of the Pacific Conference Series Vol. 230, Galaxy
  Disks and Disk Galaxies. pp 87--88

\bibitem[\protect\citeauthoryear{{Riess} et~al.,}{{Riess}
  et~al.}{1998}]{Riess_1998}
{Riess} A.~G.,  et~al., 1998, \mn@doi [AJ] {10.1086/300499}, \href
  {http://adsabs.harvard.edu/abs/1998AJ....116.1009R} {116, 1009}

\bibitem[\protect\citeauthoryear{{Rubin}, {Ford}  \& {Thonnard}}{{Rubin}
  et~al.}{1980}]{Rubin_1980}
{Rubin} V.~C.,  {Ford} Jr. W.~K.,   {Thonnard} N.,  1980, \mn@doi [ApJ]
  {10.1086/158003}, \href {http://adsabs.harvard.edu/abs/1980ApJ...238..471R}
  {238, 471}

\bibitem[\protect\citeauthoryear{{Salem}, {Besla}, {Bryan}, {Putman}, {van der
  Marel}  \& {Tonnesen}}{{Salem} et~al.}{2015}]{Salem_2015}
{Salem} M.,  {Besla} G.,  {Bryan} G.,  {Putman} M.,  {van der Marel} R.~P.,
  {Tonnesen} S.,  2015, \mn@doi [ApJ] {10.1088/0004-637X/815/1/77}, \href
  {http://adsabs.harvard.edu/abs/2015ApJ...815...77S} {815, 77}

\bibitem[\protect\citeauthoryear{{Salucci}, {Lapi}, {Tonini}, {Gentile},
  {Yegorova}  \& {Klein}}{{Salucci} et~al.}{2007}]{Salucci_2007}
{Salucci} P.,  {Lapi} A.,  {Tonini} C.,  {Gentile} G.,  {Yegorova} I.,
  {Klein} U.,  2007, \mn@doi [MNRAS] {10.1111/j.1365-2966.2007.11696.x}, \href
  {http://adsabs.harvard.edu/abs/2007MNRAS.378...41S} {378, 41}

\bibitem[\protect\citeauthoryear{{Sanders}}{{Sanders}}{2003}]{Sanders_2003}
{Sanders} R.~H.,  2003, \mn@doi [MNRAS] {10.1046/j.1365-8711.2003.06596.x},
  \href {http://adsabs.harvard.edu/abs/2003MNRAS.342..901S} {342, 901}

\bibitem[\protect\citeauthoryear{{Smith} et~al.,}{{Smith}
  et~al.}{2007}]{Smith_2007}
{Smith} M.~C.,  et~al., 2007, \mn@doi [MNRAS]
  {10.1111/j.1365-2966.2007.11964.x}, \href
  {http://adsabs.harvard.edu/abs/2007MNRAS.379..755S} {379, 755}

\bibitem[\protect\citeauthoryear{{Smolin}}{{Smolin}}{2017}]{Smolin_2017}
{Smolin} L.,  2017, \mn@doi [Physical Review D] {10.1103/PhysRevD.96.083523},
  \href {http://adsabs.harvard.edu/abs/2017PhRvD..96h3523S} {96, 083523}

\bibitem[\protect\citeauthoryear{{Snaith}, {Haywood}, {Di Matteo}, {Lehnert},
  {Combes}, {Katz}  \& {G{\'o}mez}}{{Snaith} et~al.}{2014}]{Snaith_2014}
{Snaith} O.~N.,  {Haywood} M.,  {Di Matteo} P.,  {Lehnert} M.~D.,  {Combes} F.,
   {Katz} D.,   {G{\'o}mez} A.,  2014, \mn@doi [ApJL]
  {10.1088/2041-8205/781/2/L31}, \href
  {http://adsabs.harvard.edu/abs/2014ApJ...781L..31S} {781, L31}

\bibitem[\protect\citeauthoryear{{Teyssier}}{{Teyssier}}{2002}]{Teyssier_2002}
{Teyssier} R.,  2002, \mn@doi [A\&A] {10.1051/0004-6361:20011817}, \href
  {http://adsabs.harvard.edu/abs/2002A%26A...385..337T} {385, 337}

\bibitem[\protect\citeauthoryear{{Thomas}, {Famaey}, {Ibata}, {L{\"u}ghausen}
  \& {Kroupa}}{{Thomas} et~al.}{2017}]{Thomas_2017}
{Thomas} G.~F.,  {Famaey} B.,  {Ibata} R.,  {L{\"u}ghausen} F.,   {Kroupa} P.,
  2017, \mn@doi [A\&A] {10.1051/0004-6361/201730531}, \href
  {http://adsabs.harvard.edu/abs/2017A%26A...603A..65T} {603, A65}

\bibitem[\protect\citeauthoryear{{Udry} et~al.,}{{Udry}
  et~al.}{1997}]{Udry_1997}
{Udry} S.,  et~al., 1997, in {Bonnet} R.~M.,  et~al., eds,  ESA Special
  Publication Vol. 402, Hipparcos - Venice '97. pp 693--698

\bibitem[\protect\citeauthoryear{{Verlinde}}{{Verlinde}}{2016}]{Verlinde_2016}
{Verlinde} E.~P.,  2016, \mn@doi [SciPost Physics]
  {10.21468/SciPostPhys.2.3.016}, \href
  {http://adsabs.harvard.edu/abs/2016arXiv161102269V} {2, 16}

\bibitem[\protect\citeauthoryear{{Williams}, {Belokurov}, {Casey}  \&
  {Evans}}{{Williams} et~al.}{2017}]{Williams_2017}
{Williams} A.~A.,  {Belokurov} V.,  {Casey} A.~R.,   {Evans} N.~W.,  2017,
  \mn@doi [\mnras] {10.1093/mnras/stx508}, \href
  {http://adsabs.harvard.edu/abs/2017MNRAS.468.2359W} {468, 2359}

\bibitem[\protect\citeauthoryear{{Wu}, {Famaey}, {Gentile}, {Perets}  \&
  {Zhao}}{{Wu} et~al.}{2008}]{Wu_2008}
{Wu} X.,  {Famaey} B.,  {Gentile} G.,  {Perets} H.,   {Zhao} H.,  2008, \mn@doi
  [MNRAS] {10.1111/j.1365-2966.2008.13198.x}, \href
  {http://adsabs.harvard.edu/abs/2008MNRAS.386.2199W} {386, 2199}

\bibitem[\protect\citeauthoryear{{Zhao} \& {Famaey}}{{Zhao} \&
  {Famaey}}{2012}]{Zhao_2012}
{Zhao} H.,  {Famaey} B.,  2012, \mn@doi [Physical Review D]
  {10.1103/PhysRevD.86.067301}, \href
  {http://adsabs.harvard.edu/abs/2012PhRvD..86f7301Z} {86, 067301}

\bibitem[\protect\citeauthoryear{{Zhao}, {Famaey}, {L{\"u}ghausen}  \&
  {Kroupa}}{{Zhao} et~al.}{2013}]{Zhao_2013}
{Zhao} H.,  {Famaey} B.,  {L{\"u}ghausen} F.,   {Kroupa} P.,  2013, \mn@doi
  [A\&A] {10.1051/0004-6361/201321879}, \href
  {http://adsabs.harvard.edu/abs/2013A\%26A...557L...3Z} {557, L3}

\bibitem[\protect\citeauthoryear{{de Blok} \& {McGaugh}}{{de Blok} \&
  {McGaugh}}{1998}]{Blok_1998}
{de Blok} W.~J.~G.,  {McGaugh} S.~S.,  1998, \mn@doi [ApJ] {10.1086/306390},
  \href {http://adsabs.harvard.edu/abs/1998ApJ...508..132D} {508, 132}

\bibitem[\protect\citeauthoryear{{den Heijer} et~al.,}{{den Heijer}
  et~al.}{2015}]{den_Heijer_2015}
{den Heijer} M.,  et~al., 2015, \mn@doi [A\&A] {10.1051/0004-6361/201526879},
  \href {http://adsabs.harvard.edu/abs/2015A%26A...581A..98D} {581, A98}

\bibitem[\protect\citeauthoryear{{van der Marel}, {Besla}, {Cox}, {Sohn}  \&
  {Anderson}}{{van der Marel} et~al.}{2012}]{Van_der_Marel_2012}
{van der Marel} R.~P.,  {Besla} G.,  {Cox} T.~J.,  {Sohn} S.~T.,   {Anderson}
  J.,  2012, \mn@doi [ApJ] {10.1088/0004-637X/753/1/9}, \href
  {http://adsabs.harvard.edu/abs/2012ApJ...753....9V} {753, 9}

\makeatother
\end{thebibliography}

\begin{appendix}
\section{Discretised Poisson Equation For an Exponential Disk}
\label{Appendix_A}

We use a spherical polar co-ordinate system with polar angle $\theta$. For the axisymmetric cases we consider, we do not use the azimuthal angle. Thus, Equation \ref{Poisson_equation} reduces to
\begin{eqnarray}
	\nabla^2 \Phi ~=~ \frac{1}{r^2} \frac{\partial}{\partial r} \left( r^2 \frac{\partial \Phi}{\partial r} \right) + \frac{1}{r^2 \sin \theta} \frac{\partial}{\partial \theta} \left( \sin \theta \frac{\partial \Phi}{\partial \theta} \right) \, .
	\label{Poisson_equation_spherical}
\end{eqnarray}

To solve this numerically, we discretise $r$ and $\theta$ using the integer indices $i$ and $j$, respectively. We cover the range $\theta = \left(0, \frac{\pi}{2} \right)$ using 801 equally spaced points. Radially, we use a linear grid with 801 steps from $r = 0$ out to 1 and then an exponential grid (constant $\frac{r_{i+1}}{r_i}$) such that the point at $r = 1$ is halfway between the immediately adjacent radial points, minimising sudden changes in resolution. An exponential grid at long range allows us to efficiently cover the region out to $r = 6205$ disk scale lengths, beyond which the disk potential should be very similar to that from a point mass.

The discretised version of the Laplacian operator at the cell $\left(i, j \right)$ corresponding to $r = r_i,~\theta = \theta_j$ is
\begin{eqnarray}
	\label{Discretised_Laplacian}
	\nabla^2 \Phi &=& \frac{3}{{r_{_+}}^3 ~-~ {r_{_-}}^3}\left({r_{_+}}^2 \bm{n}_{r,{_+}} ~-~ {r_{_-}}^2 \bm{n}_{r,{_-}} \right) \\
	&+& \frac{\sin \theta_{_+} \bm{n}_{\theta,{_{_+}}} ~-~ \sin \theta_{{_-}} \bm{n}_{\theta,{_-}}}{r^2 \left( \cos \theta_{_-} ~-~ \cos \theta_{_+} \right)}   ~~\text{, where} \nonumber \\
	\bm{n}_{r,_\pm} &\equiv & \frac{\Phi_{i\pm 1} ~-~ \Phi_0}{r_{i\pm 1} ~-~ r_i} \\
	\bm{n}_{\theta,_\pm} &\equiv & \pm \frac{\Phi_{j\pm 1} ~-~ \Phi_0}{r~d\theta} ~~\text{ and} \\
	r_{_\pm} &\equiv & \frac{r_i ~+~ r_{i \pm 1}}{2} \\
	\theta_\pm &\equiv & \frac{\theta_j ~+~ \theta_{j \pm 1}}{2} \, .
\end{eqnarray}

We use the convention that we only specify indices not equal to their values at the point of evaluation, where $\Phi \equiv \Phi_0$. As an example, $\Phi_{j+1} \equiv \Phi_{i,j+1}$. For the on-axis cells with $\theta = 0$, we use $\theta_- = 0$ and assume $\frac{\partial \Phi}{\partial \theta} = 0$ there due to axisymmetry. For disk cells with $\theta = \frac{\pi}{2}$, we use $\theta_{+} = \frac{\pi}{2}$ and assume that
\begin{eqnarray}
	\frac{\partial \Phi}{\partial \theta} ~=~ r\rm{e}^{-r} ~~\left(\theta = \frac{\pi}{2} \right) \, .
\end{eqnarray}

This imposes the surface density $\frac{e^{-r}}{2\pi}$, thus creating an exponential disk with scale length and $GM = 1$. As all the matter is in the disk, we need to solve the Laplace equation
\begin{eqnarray}
	\nabla^2 \Phi ~=~ 0 \, .
	\label{Laplace_equation}
\end{eqnarray}

The boundary condition in the disk plane imposes our desired surface density distribution. At the outermost radial shell, we expect $\Phi \approx -\frac{1}{r}$ but with a small correction due to the matter distribution having a finite extent. Taking this into account at the lowest order, we impose the boundary condition that
\begin{eqnarray}
	\Phi ~=~ -\frac{1}{r} \left( 1 + \frac{3 \left( 3\sin^2 \theta - 2\right)}{2r^2} \right) ~~\text{on outer shell.}
\end{eqnarray}

The potential is not updated at this radial shell. To accelerate the convergence of our algorithm, we set
\begin{eqnarray}
	\Phi \left(r = 0 \right) ~=~ -1 \, .
\end{eqnarray}

We can use Equation \ref{Discretised_Laplacian} to determine the sensitivity of $\nabla^2\Phi$ to $\Phi_0$ and thus the value $\Phi_{new}$ such that if $\Phi_0 = \Phi_{new}$, then $\nabla^2 \Phi = 0$ locally. We then apply an over-relaxation method by setting
\begin{eqnarray}
	\Phi_0 ~\to~ W_{new}\Phi_{new} ~+~ \left(1 - W_{new} \right) \Phi_0 \, .
\end{eqnarray}

Given $n_r$ radial elements, we find that it is best to start with
\begin{eqnarray}
	W_{new} ~\approx~ \frac{2}{{1 + \frac{\pi}{n_r}}} \, .
\end{eqnarray}

To get our algorithm to converge, we need to gradually reduce $W_{new}$ as it proceeds to completion, which we let it do by considering various diagnostics. We find that it is important to use $W_{new} > 1.8$ because lower values lead to extremely slow progress. However, values above 2 do not work either because they lead to instability.

We apply a stringent convergence criterion to ensure that Equation \ref{Laplace_equation} is satisfied everywhere subject to the boundary conditions. Equation \ref{Discretised_Laplacian} is used to determine whether a cell has `converged', by which we mean that $\left| \nabla^2 \Phi \right| < 10^{-4}$. On a particular grid update\footnote{i.e. going through all 8 colours of our colouring scheme}, we require convergence of all cells that are being updated (which excludes the boundary radially and the origin but not cells in the disk). The algorithm then needs to achieve this a second consecutive time, at which point we assume the solution is acceptable.

Equation \ref{Discretised_Laplacian} implies that very tiny errors in $\Phi$ lead to large errors in $\nabla^2 \Phi$ at points close to the origin. This can prevent the algorithm converging while it tries to work out the potential to e.g. the twelfth decimal place, which computers struggle to do. Thus, we increase the tolerance on $\left| \nabla^2 \Phi \right|$ by a factor of $\left(r \sin \theta \right) ^{-1}$ if this indeed raises the tolerance.

\section{Discretised Source for the QUMOND Poisson Equation}
\label{Appendix_B}

The divergence of the true gravitational field $\bm{g}$ is $\nabla \cdot \left(\nu \bm{g_{_N}} \right)$, which we find using the same discretisation scheme as in Appendix \ref{Appendix_A}. Because we need to determine $\nu$, all components of $\bm{g_{_N}}$ are required at all points surrounding the one we are considering. We already have the `divergence parts' of $\bm{g_{_N}}$ e.g. we know $\bm{n}_{r,\pm}$ at the points $\left(r_\pm, \theta \right)$. To get the other component of $\bm{g_{_N}}$ at these points, we use a centred differencing scheme.
\begin{eqnarray}
	\frac{4d\theta}{r_{_\pm}} \left. \frac{\partial \Phi}{\partial \theta}\right|_{r_{_\pm}} = \frac{\Phi_{i \pm 1, j + 1} - \Phi_{i \pm 1, j - 1}}{r_{_{i \pm 1}}} + \frac{\Phi_{i, j + 1} - \Phi_{i, j - 1}}{r_{_i}} \, .
\end{eqnarray}

A similar procedure would be inaccurate for $\frac{\partial \Phi}{\partial r}$ at $\left(r, \theta_\pm \right)$ due to our non-linear radial resolution scheme. Thus, we estimate $\frac{\partial \Phi}{\partial r}$ at $\left(r, \theta_j \right)$ based on fitting a parabola through $\Phi_{i - 1}$, $\Phi_0$ and $\Phi_{i + 1}$. After applying a similar procedure to get $\frac{\partial \Phi}{\partial r}$ at $\left(r, \theta_{j \pm 1} \right)$, we average the results to estimate $\frac{\partial \Phi}{\partial r}$ at $\left(r, \theta_\pm \right)$.

After obtaining all components of $\bm{g_{_N}}$ in this way, it is simple to multiply each potential derivative appearing in Equation \ref{Discretised_Laplacian} by the value of $\nu$ at the corresponding position, thus yielding the divergence of the true gravitational field.
\end{appendix}

\bsp
\label{lastpage}
\end{document}